\documentclass[a4paper,11pt]{article}
\usepackage{jcappub,amsmath, mathtools,ascmac,subcaption,fancybox} 

\newcommand{\nn}{\nonumber \\}
\newcommand*{\calA}{{\cal A}}
\newcommand*{\calB}{{\cal B}}

\newcommand*{\calE}{{\cal E}}
\newcommand*{\calF}{{\cal F}}

\thisfancyput(12.0cm,0cm){{RUP-24-21, YITP-24-157}}
\title{Soliton self-gravity and core-halo relation in fuzzy dark matter halos}

\author[a]{Yusuke Manita}
\author[b]{Takuya Takahashi}
\author[c,d]{Atsushi Taruya}

\affiliation[a]{Perimeter Institute for Theoretical Physics, Waterloo, Ontario, N2L 2Y5 Canada}
\affiliation[b]{Department of Physics, Rikkyo University, Toshima, Tokyo 171-8501, Japan}
\affiliation[c]{Center for Gravitational Physics and Quantum Information, Yukawa Institute for Theoretical Physics, Kyoto University, Kyoto 606-8502, Japan}
\affiliation[d]{Kavli Institute for the Physics and Mathematics of the Universe (WPI), UTIAS, The University of Tokyo, Kashiwa, Chiba 277-8583, Japan}

\emailAdd{ymanita@perimeterinstitute.ca}

\abstract{
     Fuzzy dark matter (FDM) is an attractive dark matter candidate composed of ultralight particles.  In this paper, toward a clear understanding of the core-halo relation in the FDM halos, we consider a simple model of the soliton-halo system, in which the self-gravitating soliton core is formed in the presence of Navarro–Frenk–White (NFW) halo potential as an external field. Solving numerically the Schr\"odinger-Poisson equation, the self-gravitating soliton is obtained as a ground-state solution, which is characterized by the two key parameters, i.e., size of soliton core and its strength of self-gravity relative to those of the NFW halo. Using our soliton-halo model, we investigate the properties of soliton cores found in cosmological simulation, and the key parameters characterizing these solitons are reconstructed in a self-consistent manner. Results suggest that (1) the soliton core properties depend critically on both the self-gravity of the soliton and the external potential of the host halo, and (2) the scatter observed in the core-halo relation cannot be explained solely by the one in the halo's concentration-mass relation, as previously suggested, but also significantly influenced by intrinsic features of the soliton core, potentially arising from local dynamics at the halo center. We also demonstrate that the FDM mass can be reconstructed from the simulation data characterizing the halo density profile, providing a basis for  applying the model to observational studies.
}

\begin{document}
\maketitle
\flushbottom

\section{Introduction}

The dark matter is the invisible matter component that makes up $\sim30\%$ of the energy contents of the universe. Despite the numerous attempts to detect dark matter directly, its nature remains unknown. In the standard cosmological model, the dark matter is described by the particle with negligible velocity dispersion, referred to as the cold dark matter (CDM). While the CDM paradigm has been successful in explaining the observation on scales larger than galaxies~\cite{Chabanier:2019eai,SDSS:2003tbn}, it faces the problem of discrepancy from the observations at small scales (see e.g., refs~\cite{Bullock:2017xww,Tulin:2017ara} for review). 

One of the attractive alternatives to the CDM is the dark matter described by the ultralight scalar particle with mass $m_\psi\sim10^{-22}$ eV, often called fuzzy dark matter (FDM)~\cite{Hu:2000ke}.
Since the FDM has a galaxy scale long de Broglie wavelength 
\begin{align}
    \lambda_{\rm dB}\simeq 0.48\ {\rm kpc}\left(\frac{10^{-22}{\rm eV}}{m_{\rm \psi}}\right)\left(\frac{250{\rm km \ s^{-1}}}{v}\right)\,,
\end{align}
the wave-like behavior appears at astrophysical scales. It suppresses the small-scale structure, while remains unchanged in the evolution and formation of the large scale structure~\cite{Niemeyer:2019aqm,Hui:2021tkt,Ferreira:2020fam,Hui:2016ltb}. In addition, the FDM has strong motivation from the theory beyond the Standard Model of particle physics, as exemplified by axion-like particles predicted by string theory~\cite{Marsh:2015xka,Svrcek:2006yi,Arvanitaki:2009fg}. Therefore, investigating the properties of FDM and observations at small scale can provide opportunities to explore the nature of the dark matter related to unknown physics.

The dynamics and the structure formation of FDM are governed by the Schr\"{o}dinger equation coupled to the Poisson equation. 
Numerical simulations of this system have shown that there appear several unique features in the structure of dark matter halos~\cite{Schive:2014dra,Schive:2014hza}. The density profile has a flat core called {\it soliton core} at the central region, with the outer regions resembling those found in a CDM halo, which are well described by the Navarro-Frenk-White (NFW) profile ~\cite{Navarro:1995iw,Navarro:1996gj}.
In particular, early numerical works suggest the universal relation between the configuration of the soliton core and the host halo. It is expressed as the pawer-law behavior of the radius and the mass of the soliton core with respect to the halo mass~\cite{Schive:2014hza,Nori:2020jzx,Mocz:2017wlg,Nori:2022afw}. However, the core-halo relation of the numerical simulations in different setups does not agree~\cite{Chan:2021bja}.  In addition, there is a large scatter in the core-halo relation even in the same setup of the each simulation, but what cause of the diversity and the scatter has not yet been clarified.

So far, most of the studies has been based on numerical simulations~\cite{Schwabe:2016rze,Du:2016aik,Mocz:2017wlg,Chan:2021bja,Mina:2020eik,Nori:2020jzx,Nori:2022afw,Elgamal:2023yzt,May:2021wwp,Veltmaat:2018dfz,Veltmaat:2019hou}, and little analytical work has been done (but see~\cite{Chavanis:2019faf,Desjacques:2019zhf,Kawai:2023okm}). In order to elucidate the diversity in numerical simulations as well as to understand the physical properties of the core-halo structure, analytical treatment would be helpful. The difficulty in treating the Schr\"{o}dinger-Poisson system analytically comes from the nonlinearity originated from the self-gravity, and this is also the case even if we consider a stationary spherical system. In ref.~\cite{Taruya:2022zmt}, the authors, including one of the authors of this paper, presented an analytical description of the soliton core structure, and derived the core-halo relation, ignoring the soliton self-gravity but taking the influence from the gravity sourced by the host halo into account. With this simplification, Schr\"{o}dinger-Poisson equation is reduced to a linear system, and they succeeded to obtain accurate analytical solutions for the soliton core, which is described by the ground-state wavefunction of the linear Schrodinger equation. As a result, ref.~\cite{Taruya:2022zmt} partly reproduces the core-halo relation obtained by numerical simulations. Interestingly, the relation they derived involves an additional factor which crucially depends on the halo concentration. Since the concentration-mass relation is known to exhibit a large scatter in the CDM simulations~\cite{Bullock:1999he}, the authors of ref.~\cite{Taruya:2022zmt} suggest that the diversity of the core-halo relation is originated from the scatter in the  concentration-mass relation (see ref.~\cite{Kawai:2023okm} for more quantitative discussion).

In this paper, extending the work by ref.~\cite{Taruya:2022zmt}, we further investigate the soliton core structure and its relation to host halo properties from a simplified setup. In particular, taking the soliton's self-gravity into account, we critically examineits impact on the the core-halo relation. We consider the FDM halo as a superposition of wavefunctions of the ground-state representing the soliton core and excited states representing the host halo, the latter of which are treated as an external background, and is described by the NFW density profile. Then, the soliton core, described as a ground-state wave function, is characterized by the two key parameters. While one of them is related to the size of the soliton core, another characterizes its strength of self-gravity. Both of them are dimensionless parameters relative to the halo characteristics. In this respect, our soliton-halo system allows us to discuss the core-halo relation more consistently. 

Solving numerically the Schr\"odinger-Poisson equation, the soliton solution is obtained, and comparing it with those found in numerical simulations, the key parameters of the soliton core properties are reconstructed for each halo. The results indicate that most of the FDM halos observed in the simulations lies at the parameter region where both the self-gravity of the soliton core and the host halo are important. Further, the parameter which controls the soliton self-gravity exhibits a large scatter among host halos having the same mass. This seems difficult to explain solely based on the intrinsic properties of halos arising from the concentration-mass relation, in contrast to what has been suggested by refs.~\cite{Taruya:2022zmt,Kawai:2023okm}. Rather, our results imply that the core-halo relation and its diversity found in numerical simulations are originated not only from the host halo properties, associated with the formation and merger histories, but also from the soliton intrinsic features, perhaps arising from the local dynamics around halo center. 

This paper is organized as follows. In section~\ref{sec:formulation}, we introduce a model of the soliton core-halo system, in which the self-gravity of the soliton is properly taken into account under the external halo potential. This system is reduced to a stationary problem of the spherically symmetric Schrodinger-Poisson equation, and we show in section~\ref{sec:two_limiting_cases} that it involves the two limiting cases, in which several properties of the ground-state wavefunction are analytically derived. In section~\ref{sec:numerical_results}, we present the numerical method to solve the introduced Schr\"{o}dinger-Poisson system, and show the results of the numerical computation. In section~\ref{sec:parameter_reconstruction}, the model parameters and the FDM mass are reconstructed from the data of the FDM halo simulations. We discuss the implication of the result of the reconstruction to the core-halo relation. Section~\ref{sec:summary} is devoted for summary and discussion. For the whole of this paper, we use the notation $c=\hbar=1$.

\section{Schr\"{o}dinger-Poisson system}
\label{sec:formulation}

In this section, we introduce an analytical model of the FDM halo. We here write down the Schr\"{o}dinger-Poisson equations as the basic equations describing the FDM halos. In the comoving coordinate system, they are given by (e.g., ~\cite{Marsh:2015xka})
\begin{align}
    i\frac{\partial \psi}{\partial t}&=\left[-\frac{1}{2m_{\psi}a^2}\nabla^2+m_{\psi}\Phi\right]\psi\,,\label{eq:schrodinger}
    \\
    \nabla^2 \Phi&=\frac{4\pi G}{a}\rho_\psi\,,\label{eq:poisson}
\end{align}
where $\psi$ is the wavefunction of FDM with the mass $m_\psi$. Here, $a$ is a scale factor, and $\Phi$ is the gravitational potential sourced by the density of FDM
\begin{align}
    \rho_\psi :=m_\psi|\psi|^2\,. \label{eq:density}
\end{align}

\subsection{A model of core-halo system}
FDM halo simulations based on the Schr\"{o}dinger-Poisson equation have confirmed the existence of a core-like density profile structure at the center of the halo, known as a soliton core, which has a de Broglie wavelength scale~\cite{Schive:2014dra,Schive:2014hza,May:2021wwp,Chan:2021bja}. The soliton core fits well with the ground-state of the stationary spherically symmetric solution of the Schr\"{o}dinger-Poisson equation. Outside the soliton core, there exist density fluctuation granules of comparable size to the soliton core~\cite{Schive:2014dra,Schive:2014hza}, which are caused by interference of excited states~\cite{Yavetz:2021pbc}. There are several framework to deal with such a granule structure of the stationary halo based on Schr\"odinger-Poisson equation (e.g.,~\cite{Li:2020ryg,Zagorac:2021qxq,Lin:2018whl,Yavetz:2021pbc}). To describe the soliton-like core structure at the halo center, one simple approach is to time-average the contribution from these granules, which are represented as a superposition of excited states with non-zero angular momentum. Then, the density profile outside the soliton core is close to the Navarro-Frenk-White (NFW) profile, given  by~\cite{Navarro:1995iw,Navarro:1996gj}. 
\begin{align}
    \rho_{\rm NFW}(r)=\frac{\rho_{\rm s}}{(r/r_{\rm s})(1+r/r_{\rm s})^2}\,,
    \label{eq:rho_NFW}
\end{align}
where $r$ represents the radial distance from the halo center, and $\rho_{\rm s}$ and $r_{\rm s}$ are constant parameters with the dimensions of density and length.The overall density profile of the FDM halo is thus represented as the sum of the soliton core and the NFW density profile:
\begin{align}
    \rho_{\psi}(r)\approx\rho_{\rm sol}(r)+\rho_{\rm NFW}(r).
    \label{eq:timeaverage}
\end{align}
Since the Poisson equation is linear, the potential $\Phi$ is the sum of those originating from $\rho_{\rm sol}$ and $\rho_{\rm NFW}$ as
\begin{align}
    \Phi(r)=\Phi_{\rm sol}(r)+\Phi_{\rm NFW}(r)\,,
\end{align}
where $\Phi_{\rm sol}$ and $\Phi_{\rm NFW}$ are the solutions of each Poisson equation
\begin{align}
    \frac{1}{r^2}\frac{d}{d r}\left(r^2\frac{d \Phi_{\rm sol}}{d r}\right)&=\frac{4\pi G}{a} \rho_{\rm sol}\,,
    \\
    \frac{1}{r^2}\frac{d}{d r}\left(r^2\frac{d \Phi_{\rm NFW}}{d r}\right)&=\frac{4\pi G}{a} \rho_{\rm NFW}\,.
\end{align}
The Poisson equation for the NFW density profile can be solved as
\begin{align}
    \Phi_{\rm NFW}(r)=-4\pi G\frac{\rho_{\rm s} r_{\rm s}^2}{a}\frac{\log(1+r/r_{\rm s})}{r/r_{\rm s}}\,.
\end{align}
We assume the stationary spherical wave function, i.e., $\psi=u(r)e^{-iE\tau}$ with $\tau=\int^tdt'/\{a(t')\}^2$.
In this setup, the Schr\"{o}dinger-Poisson equation can be reduced to the following form:
\begin{align}
    \frac{1}{2m_\psi r^2}\frac{d}{d r}\left(r^2\frac{d u}{d r}\right)
    -m_\psi\left[a^2\Phi_{\rm sol}-4\pi G\rho_{\rm s} r_{\rm s}^2 a\frac{\log(1+r/r_{\rm s})}{r/r_{\rm s}}-\frac{E}{m_{\psi}}\right]u&=0\,, \label{eq:schrodingerl0} 
    \\
    \frac{1}{r^2}\frac{d}{d r}\left(r^2\frac{d \Phi_{\rm sol}}{d r}\right)-\frac{4\pi G}{a}m_\psi u^2&=0\,. 
    \label{eq:poissonl0}
\end{align}
The previous work~\cite{Taruya:2022zmt} develops an analytical method to investigate the particular case of this equation while neglecting the contribution of soliton self-gravity\footnote{To be precise, ref.~\cite{Taruya:2022zmt} ignores the soliton self-gravity in their  analytical model, but estimate its impact on the soliton core structure based on perturbative treatment. They found that the self-gravity of the soliton is not entirely negligible. }.

\subsection{Dimensionless equations}

To clarify the structure of the Schr\"{o}dinger-Poisson equation, it is convenient to introduce dimensionless quantities as
\begin{align}
    \tilde{u}&:=\sqrt{8\pi Gm_\psi^3r_{\rm s}^4 a}u\,,\\
    \tilde{\Phi}&:=2m_\psi^2r_{\rm s}^2a^2 \Phi_{\rm sol}\,, \\
    x&:=\frac{r}{r_{\rm s}}\,, \\
    \calE&:=2m_\psi r_{\rm s}^2 E\,.
\end{align}
We also define a dimensionless parameter
\begin{align}
    \alpha&:=8\pi Gm_\psi^2\rho_{\rm s} r_{\rm s}^4a\,.
    \label{eq:def_alpha}
\end{align}
By applying these variables in eq.~\eqref{eq:schrodingerl0} and eq.~\eqref{eq:poissonl0}, the dimensionless Schr\"{o}dinger-Poisson equation is given by  
\begin{align}
    \frac{d^2\tilde{u}}{dx^2}+\frac{2}{x}\frac{d\tilde{u}}{dx}
    &=\left[-\calE+\tilde{\Phi}-\alpha\frac{\log(1+x)}{x}\right]\tilde{u}\,,
    \label{eq:schrodinger2}
    \\
    \frac{d^2\tilde{\Phi}}{dx^2}+\frac{2}{x}\frac{d\tilde{\Phi}}{dx}
    &=\tilde{u}^2 \,.
    \label{eq:poisson2}
\end{align}
Since we are interested in the bound state solution of the FDM as mentioned above, we focus on the solution with the regularity at the center and a decaying boundary condition at infinity.
In this case, the eigenvalue ${\cal E}$ is discretized.
Due to the non-linearity of the self-gravity, the solutions depend on the amplitude of the wavefunction.
Here, we introduce a parameter that characterizes the strength of the self-gravity as
\begin{align}
    \beta\coloneqq \tilde{u}(0)^2 \,.
    \label{eq:defbeta}
\end{align}
Therefore, the boundary conditions can be formally written as
\begin{align}
    \tilde{u}(0)&=\sqrt{\beta}\,,\label{eq:boundarycondition1}\\
    \tilde{u}'(0)&=0\,,\\
    \tilde{\Phi}'(0)&=0\,,\\
    \tilde{u}(\infty)&=0\,,\label{eq:boundarycondition4}\\
    \tilde{\Phi}(\infty)&=0\,.\label{eq:boundarycondition5}
\end{align}
In particular, the central density of the soliton core, $\rho_c:=m_{\psi}u(0)^2$, is expressed as
\begin{align}
    \rho_c
    =\frac{\beta}{8\pi G m_\psi^2 r_{\rm s}^4 a}
    =\frac{\beta}{\alpha}\,\rho_{\rm s}\,.
    \label{eq:coredensity}
\end{align}

\section{Analytical properties of soliton-halo system}
\label{sec:two_limiting_cases}

The setup described in the previous section involves the two dimensionless parameters, $\alpha$ and $\beta$. Depending on the choice of their parameters, the system can be reduced to a more simplified form. In the following, we consider the two limiting cases, i.e., the case where either the background halo potential or the soliton self-gravity is ignored, summarized in Table~\ref{tab:summary}. We derive several key quantities characterizing the conditions for two limiting cases. 

\subsection{Two limiting cases}

\subsubsection{Limit A: the case ignoring the external halo potential}
\label{subsubsec:Limit_A}

Let us first consider the case where the potential of the host halo, described by the NFW profile, is ignored. The Schr\"{o}dinger-Poisson equations eq.~\eqref{eq:schrodinger2} and eq.~\eqref{eq:poisson2} are reduced to
\begin{align}
    \frac{1}{x^2}\frac{d}{d x}\left(x^2\frac{d\tilde{u}}{d x}\right)&=\left(-\calE+\tilde{\Phi}\right)\tilde{u}\,,
    \label{eq:schrodingerSG}\\
    \frac{1}{x^2}\frac{d}{d x}\left(x^2\frac{d\tilde{\Phi}}{dx}\right)&= \tilde{u}^2\,.
    \label{eq:poissonSG}
\end{align}
This set of equations is known to be invariant under
\begin{align}
   (x, \tilde{u}, \tilde{\Phi}, \calE) \to (\lambda^{-1} x, \lambda^{2} \tilde{u}, \lambda^{2} \tilde{\Phi}, \lambda^{2} \calE)\,,
    \label{eq:sca}
\end{align}
where the parameter $\lambda$ beging a positive real constant~\cite{Schive:2014hza,Moroz:1998dh}. Note that this scaling symmetry is broken when the NFW potential is included beyond the limit A.
When $\beta$ is set to 1, i.e., $\tilde{u}(0)=1$, the ground-state solution can be obtained numerically as $\calE\simeq-0.979$ and $\tilde{\Phi}(0)\simeq-1.90$~\cite{Marsh:2015wka}. Thanks to the scaling symmetry, we can construct the solution scaled by an arbitrary 
constant $\beta$ as
\begin{align}
    \tilde{\Phi}(0)&\simeq-1.90\,\beta^{1/2}\,,
    \label{eq:phinoSG}
    \\
    {\cal E}&\simeq-0.979\,\beta^{1/2}\,.
    \label{eq:EnoSG}
\end{align}
Also, the dimensionless core radius $x_{\rm c}$,
defined as the radius at which the soliton density is half of its central value, i.e,
\begin{align}
    \tilde{u}(x_{\rm c})^2 \equiv \frac{1}{2}\tilde{u}(0)^2\,,
    \label{eq:def_core_radius}
\end{align}
is also obtained, and for $\beta = 1$, the numerical calculations yield $x_{\rm c} \simeq 1.55 (\equiv q)$. 
Putting $\beta$ back to the expression with the scaling law, we thus have
\begin{align}
    x_{\rm c} = q\beta^{-1/4}\,.
    \label{eq:coreradius}
\end{align}

Recall that the radius $r_{\rm c}$ is related to the dimensionless quantity $x_{\rm c}$ through $r_{\rm c}=x_{\rm c}\,r_{\rm s}$, 
combining eq.~\eqref{eq:coredensity} with eq.~\eqref{eq:coreradius} leads to the following relation~\cite{Schive:2014hza}:
\begin{align}
    \rho_{\rm c} &\simeq 
    \frac{q^4}{8\pi G m_\psi^2a r_{\rm c}^4}
    \simeq\frac{1.94}{a}\left[M_{\odot} \mathrm{pc}^{-3}\right]\left(\frac{m_\psi}{10^{-23}{\rm eV}}\right)^{-2}\left(\frac{r_{\mathrm{c}}}{1 \mathrm{kpc}}\right)^{-4}\,,
    \label{eq:centralcoredensity}
\end{align}
where $M_{\odot}$ and $\mathrm{pc}^{-3}$ denotes solar mass and parsection

\begin{table}[tbp]
    \centering
    \begin{tabular}{c|ccc}
    \hline
     & Dominant potential & Parameter region & Core radius $x_{\rm c}$ \\ \hline
    Limit A & Soliton core & $\beta\gg\beta_{\rm crit}(\alpha)$ & $q\beta^{-1/4}$ \\
    Limit B & Host halo & $\beta\ll\beta_{\rm crit}(\alpha)$ & $\bar{x}_c(\alpha)\approx\bar{p}\alpha^{-1/3}$ \\ \hline
\end{tabular}
    \caption{Summary of the two limiting cases.}
    \label{tab:summary}
\end{table}

\subsubsection{Limit B: the case ignoring the soliton's self-gravity potential}
\label{subsubsec:Limit_B}

Next consider another limiting case, in which we ignore the contribution of the soliton core to the gravitational potential. We call it  {\it limit B}. This is to drop the second term in the square bracket of  the right hand side of eq.~\eqref{eq:schrodinger2}. Then, the Schr\"odinger-Poisson equations are reduced to the linear Schr\"odinger system, and the equation to solve becomes
\begin{align}
    \frac{1}{x^2}\frac{d}{d x}\left(x^2\frac{d\tilde{u}}{d x}\right)&=\left[-\calE-\alpha\frac{\log(1+x)}{x}\right]\tilde{u}\,.
    \label{eq:TS_Schrodinger}
\end{align}
Note that this Schr\"{o}dinger equation is no longer invariant under the scaling transformation eq.~\eqref{eq:sca}. In ref.~\cite{Taruya:2022zmt}, applying the uniform asymptotic approximation, the authors obtain the analytical expressions for the eigenfunction of eq.~\eqref{eq:TS_Schrodinger}, which are found to describe very accurately the soliton solutions in the case of $\alpha\gg1$. The eigenvalues are then obtained by solving the following transcendental equation:
\begin{align}
    {\rm Ai}(z(0))=0\,,
    \label{eq:airyfunc}
\end{align}
where ${\rm Ai}(z)$ denotes the Airy function of the first kind, and $z(x)$ is defined as
\begin{align}
    z(x):=\alpha^{1/3}\left[\frac{3}{2}\int_{x}^{x_{\rm tp}}dx'\sqrt{\frac{\log(1+x')}{x'}+\frac{\calE}{\alpha}}\right]^{3/2}\,,
\end{align}
with the quantity $x_{\rm tp}$ being the turning point satisfying the relation $\log(1+x_{\rm tp})/x_{\rm tp}+\calE/\alpha=0$.
Imposing further the condition $x_{\rm tp}\ll1$, it is possible to get a more concise expression for the eigenvalues, and the ground-state eigenvalue is approximately expressed as 
\begin{align}
    \calE \approx -\alpha+\left(\frac{9\pi}{16}\alpha\right)^{2/3}\,.
    \label{eq:calEalpha}
\end{align}

In ref.~\cite{Taruya:2022zmt}, taking advantage of the analytical form of the ground-state eigenfunction, the expression of the core radius is also obtained in a simple form. Denoting its dimensionless core radius by $\bar{x}_c(\alpha)$, we have
\begin{align}
    \bar{x}_c(\alpha)\approx p\sqrt{\frac{6}{\alpha(1+\mathcal{E} / \alpha)}}\,
    \label{eq:xcbar_wkb}
\end{align}
with the quantity $p$ being the numerical constant, $p=0.65$~\cite{Taruya:2022zmt}. Substituting eq.~\eqref{eq:calEalpha} into the above, a further approximation is obtained:
\begin{align}
    \bar{x}_c(\alpha)\approx\sqrt{6}p\left(\frac{16}{9\pi\alpha}\right)^{1/3}\equiv \bar{p}\alpha^{-1/3}\,.
    \label{eq:TS_core_rad}
\end{align}
with $\bar{p}:=\sqrt{6}(16/9\pi)^{1/3}\simeq1.32$.

\subsubsection{Critical region}
\label{subsubcec:critical}

The two limiting cases described in section~\ref{subsubsec:Limit_A} and \ref{subsubsec:Limit_B} appear as asymptotic solutions, and in general, the ground-state solution of Schr\"odinger-Poisson equations eq.~\eqref{eq:schrodinger2} and eq.~\eqref{eq:poisson2} is neither the limit A nor limit B. Nevertheless, depending on the choice of parameters, one expects that the solution exhibits the behaviors similar to either the limit A or limit B. To clarify the boundary of the regions where the solution is expected to behave like the limit A or B, we introduce the critical value by equating the dimensionless core radius for limit A and B, namely, eq.~\eqref{eq:coreradius} and $\bar{x}_c(\alpha)$. Thus, for a given $\alpha$, the critical value is defined as
\begin{align}
    \beta_{\rm crit}(\alpha):=\left(\frac{q}{\bar{x}_c(\alpha)}\right)^4\,.
    \label{eq:goodapprox}
\end{align}
In particular, for the region of $\alpha\gg1$, the above expression is simplified by using eq.~\eqref{eq:TS_core_rad}as
\begin{align}
    \beta_{\rm crit}(\alpha)\approx\left(\frac{\bar{p}}{q}\right)^{-4}\alpha^{4/3}\simeq1.92\alpha^{4/3}\,.
    \label{eq:betacrit}
\end{align}
The ground-state solution is thus expected to behave like the limit A in the region of $\beta\gg\beta_{\rm crit}$, while the behavior like the limit B appears in the region of $\beta\ll\beta_{\rm crit}$.  

\begin{figure}[tbp]
    \centering
    \includegraphics[scale=0.5]{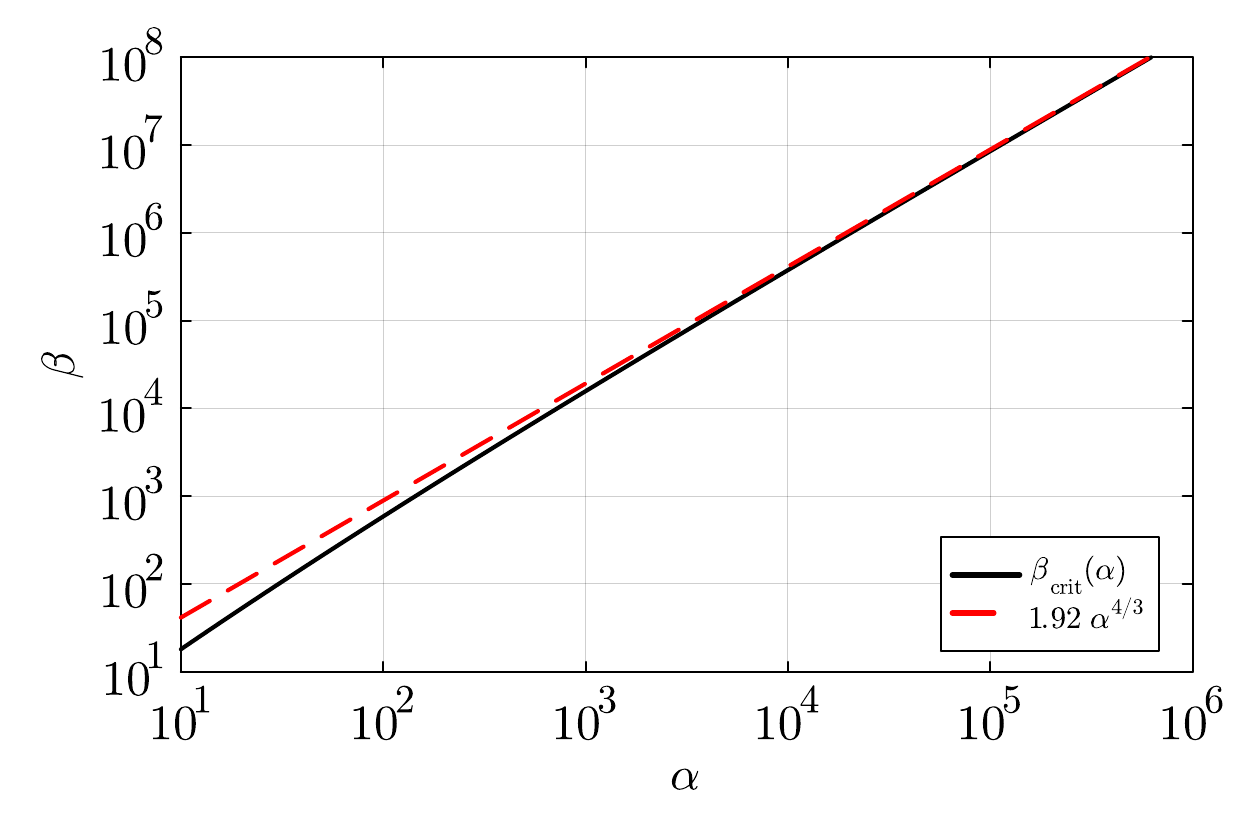}
    \caption{The black line represents the critical value of $\beta$ where the transition occurs for each $\alpha$ defined by eq.~\eqref{eq:goodapprox}. The red dashed line represents the approximated value of $\beta_{\rm crit}$ given by eq.~\eqref{eq:betacrit}.
    }
    \label{fig:alpha_beta}
\end{figure}

Figure~\ref{fig:alpha_beta} illustrates the critical value of $\beta$. The black solid line shows eq.~\eqref{eq:goodapprox} with the quantity $\bar{x}_{\rm c}(\alpha)$ being evaluated numerically by solving  eq.~\eqref{eq:airyfunc}. The red dashed line represents the result from the approximate expression of $\beta_{\rm crit}$ at eq.~\eqref{eq:betacrit}, which gives an accurate estimation of the critical value at $\alpha\gg1$.

As defined above, the asymptotic behavior of the core radius in terms of $\beta$ with $\alpha\gg1$ can be expressed as
\begin{align}
    r_{\rm c} \approx
    \begin{cases}
      q\beta^{-1/4} r_{\rm s}\,& (\text{limit A},~\beta\gg\beta_{\rm crit}) \,,
      \\
      \\
      \bar{p}\alpha^{-1/3} r_{\rm s}\,& (\text{limit B},~ \beta\ll\beta_{\rm crit})\,.
    \end{cases}
    \label{eq:rcexpression}
\end{align}

\subsection{Core mass}
Next, we discuss the core mass, which characterize the core-halo relation. We define the core mass as the enclosed total mass of dark matter within the core radius $r_{\rm c}$ as
\begin{align}
    M_{c}&:=M_{{\rm c, sol}}+M_{{\rm c, NFW}}\,;\label{eq:mctotal}\\
    M_{{\rm c, sol}}&:=4\pi\int_0^{r_{\rm c}}dr\ r^2\,\rho_{\rm sol}(r)\,,
    \label{eq:mcsol}\\
    M_{{\rm c, NFW}}&:=4\pi\int_0^{r_{\rm c}}dr\ r^2\,\rho_{\rm NFW}(r)=4\pi\rho_{\rm s} r_{\rm s}^3\left(-\frac{x_{\rm c}}{1+x_{\rm c}}+\log(1+x_{\rm c})\right)\,.
    \label{eq:mcnfw}
\end{align}
Here, we consider the contributions both from the soliton and the host halo.

In both of the limits A and B, the soliton core density $\rho_{\rm sol}(r)$ is shown to be well fitted by the formula~\cite{Schive:2014hza,Taruya:2022zmt}
\begin{align}
    \rho_{\rm sol}^{\rm (fit)}(r)=\frac{\rho_c}{(1+c(r/r_{\rm c})^2)^8}\,,
    \label{eq:fitting_formula}
\end{align}
with the constant $c$ setting to $c\simeq0.091$, so as to satisfy $\rho_{\rm sol}^{\rm (fit)}(r_{\rm c})=\rho_c/2$. Substituting this into eq.~\eqref{eq:mcsol}, the soliton core mass recast as
\begin{align}
    M_{{\rm c, sol}}& =4\pi b\rho_c r_{\rm c}^3\,,
    \label{eq:mcsol_fit}
\end{align}
where the constant $b$ is determined numerically as
\begin{align}
    b:=\int_0^1~dx\frac{x^2}{(1+c x^2)^8}\simeq 0.222\,.
\end{align}
Further, we rewrite eq.~\eqref{eq:mcsol_fit} with the expression of the dimensionless core radius given at eq.~\eqref{eq:coreradius} in the limit A and at eq.~\eqref{eq:TS_core_rad} in the limit B. We obtain 
\begin{align}
    M_{{c,\rm sol}}\approx
    \begin{dcases}
        4\pi \rho_{\rm s} r_{\rm s}^3 b q^3 \frac{\beta^{1/4}}{\alpha}\,
        & (\text{limit A},~ \beta\gg\beta_{\rm crit}),
        \\
        4\pi \rho_{\rm s} r_{\rm s}^3  b\bar{p}^3\frac{\beta}{\alpha^2}\,
        & (\text{limit B},~ \beta\ll\beta_{\rm crit}).
    \end{dcases}
    \label{eq:Mccase1}
\end{align}

Similarly, substituting the core radius from eq.~\eqref{eq:coreradius} or eq.~\eqref{eq:TS_core_rad} into eq.~\eqref{eq:mcnfw}, the core mass coming from the host halo contribution is expressed in the two limiting cases as 
\begin{align}
    M_{{c,\rm NFW}}\approx
    \begin{cases}
       {\displaystyle  4\pi\rho_{\rm s}r_{\rm s}^3\left\{-\frac{q\beta^{-1/4}}{1+q\beta^{-1/4}}+\log(1+q\beta^{-1/4})\right\}\,}
        & (\text{limit A},~ \beta\gg\beta_{\rm crit}),
        \\
        \\
        {\displaystyle 4\pi\rho_{\rm s}r_{\rm s}^3\left\{-\frac{\bar{p}\alpha^{-1/3}}{1+\bar{p}\alpha^{-1/3}}+\log(1+\bar{p}\alpha^{-1/3})\right\}\,}
        & (\text{limit B},~ \beta\ll\beta_{\rm crit}).
    \end{cases}
    \label{eq:Mccase2}
\end{align}
With the expressions given above, we can analytically evaluate which contribution dominates the total core mass.

By assuming $x_{\rm c} \ll 1$, from eqs.~\eqref{eq:mcnfw}, \eqref{eq:mcsol_fit}, and \eqref{eq:coredensity}, the ratio of the two core masses is given by
\begin{align}
    \frac{M_{c, \rm sol}}{M_{c,\rm NFW}}\approx 2b~\frac{\beta}{\alpha}x_{\rm c}\,.
    \label{eq:coreratio}
\end{align}
In the region where the limit A valid $(\beta\gg\beta_{\rm crit})$, since the core radius is given by eq.~\eqref{eq:coreradius}, the ratio of the two core masses is given by
\begin{align}
    \frac{M_{\rm c, sol}}{M_{\rm c, NFW}}\approx 2bq~\frac{\beta^{3/4}}{\alpha}\,.
\end{align}
From eq.~\eqref{eq:betacrit}, in the region where the limit A is valid $(\beta\gg\beta_{\rm crit})$, the above equation can be evaluated as $M_{\rm c, sol}/M_{\rm c, NFW}\gg1.12$.
Thus, the contribution to the core mass from the soliton density is always larger than from the host halo density. 
On the other hand, in the region where the limit B is valid $(\beta\ll\beta_{\rm crit})$, from eqs.~\eqref{eq:TS_core_rad} and \eqref{eq:coreratio}, the ratio of the two core masses is given by
\begin{align}
    \frac{M_{c, {\rm sol}}}{M_{c, {\rm NFW}}}\approx 2b\bar{p}~\frac{\beta}{\alpha^{4/3}}\,.
\end{align}
If $\beta\lesssim(2b\bar{p})^{-1}\alpha^{4/3}=1.71\alpha^{4/3}\sim\beta_{\rm crit}$, it results in $M_{{\rm c, NFW}}\gtrsim M_{{\rm c, sol}}$.
Therefore, the contribution to the core mass from the host halo is larger than from the soliton core in the most regions where the limit B is valid $(\beta\ll\beta_{\rm crit})$.

From the above, the asymptotic behavior of the total core mass for each $\alpha$ is given by
\footnote{
In the region of the limit B is valid, the Schr\"{o}dinger-Poisson equation can be approximated as a linear equation, so the amplitude of the density is indeterminate. In ref.~\cite{Taruya:2022zmt}, the amplitude of the density is determined by substituting eq.~\eqref{eq:TS_core_rad} in eq.~\eqref{eq:centralcoredensity} as
\begin{align}
    M_{\rm c}\stackrel{?}{=}4\pi\rho_{\rm s}r_{\rm s}^3\times\frac{bq^4}{\bar{p}}\alpha^{-2/3}\,.
    \label{eq:maceqquestion}
\end{align}
The formula for the central density eq.~\eqref{eq:centralcoredensity} does not hold in the region where $\beta\ll\beta_{\rm crit}$, thus the above derivation of the core mass is not appropriate. Nevertheless, this core mass approximately matches the first formula of eq.~\eqref{eq:Mccase}, except for a subtle factor difference. Assuming $x_{\rm c}\ll1$, the first formula of eq.~\eqref{eq:Mccase} can be approximated as
\begin{align}
    4\pi\rho_{\rm s}r_{\rm s}^3\times\frac{x_{\rm c}^2}{2}=4\pi\rho_{\rm s}r_{\rm s}^3\times\frac{\bar{p}^2}{2}\alpha^{-2/3}\,,
\label{eq:maceqcorrect}
\end{align}
Therefore, eqs.~\eqref{eq:maceqquestion} and \eqref{eq:maceqcorrect} match except for a factor of around 1.6.
}
\begin{align}
    M_{\rm c}(\alpha)\approx
    \begin{dcases}
        {\displaystyle 4\pi \rho_{\rm s} r_{\rm s}^3 b q^3 \frac{\beta^{1/4}}{\alpha}\,,}
        & (\text{limit A},~ \beta\gg\beta_{\rm crit})
        \\
        {\displaystyle 4\pi\rho_{\rm s}r_{\rm s}^3\left\{-\frac{\bar{p}\alpha^{-1/3}}{1+\bar{p}\alpha^{-1/3}}+\log \left(1+\bar{p}\alpha^{-1/3}\right)\right\}\,.}
        & (\text{limit B},~ \beta\ll\beta_{\rm crit})
    \end{dcases}
    \label{eq:Mccase}
\end{align}
In the limit case A, by substituting eqs.~\eqref{eq:def_alpha} and \eqref{eq:coreradius} into the first formula of eq.~\eqref{eq:Mccase}, we obtain the scaling law for the core mass
\begin{align}
    M_{{\rm c, sol}}&=\frac{5.41\times 10^{9}}{a}\left(\frac{m}{10^{-23}{\rm eV}}\right)^{-2}\left(\frac{r_{\rm c}}{1{\rm kpc}}\right)^{-1}\left[M_{\odot}\right]\,,
    \label{eq:mcsol_unit}
\end{align}

\subsection{Maximam core radius and minimam core mass}
\label{app:coremass}
In this model, we can discuss the limits of values that the quantities characterizing the solito core can take. The maximum core radius for a given $\alpha$ is given by eq.~\eqref{eq:rcexpression} as
\begin{align} 
    r_{\rm c,max}(\alpha)\approx\bar{p}\alpha^{-1/3} r_{\rm s}\,,
    \label{eq:rcmin}
\end{align}
which corresponds to the limit B. This is because the dimensionless core radius in the limit A is a decreasing function of $\beta$. On the other hand, the minimum core mass is approximately given by
\begin{align}
    M_{{\rm c, min}}(\alpha)\approx 4\pi\rho_{\rm s} r_{\rm s}^3\left(-\frac{\bar{p}\alpha^{-1/3}}{1+\bar{p}\alpha^{-1/3}}+\log \left(1+\bar{p}\alpha^{-1/3}\right)\right)\,.
    \label{eq:coremassmin}
\end{align}
This follows from the numerical verification that, assuming $\beta_{\rm crit}(\alpha)$ is expressed by equation eq.~\eqref{eq:betacrit}, the inequality
\begin{align*}
    -\frac{\bar{p}\alpha^{-1/3}}{1+\bar{p}\alpha^{-1/3}}+\log \left(1+\bar{p}\alpha^{-1/3}\right) 
    < b q^3 \frac{\beta_{\rm crit}(\alpha)^{1/4}}{\alpha}\,,
\end{align*}
holds, at least in $10^{1}<\alpha<10^{8}$. Therefore, the minimum core mass corresponds to the core mass in the region where the limit B is valid $(\beta\ll\beta_{\rm crit})$.

In the previous subsections, the maximum core radius and the minimum core mass for a given $\alpha$ are obtained as eq.~\eqref{eq:rcmin} and eq.~\eqref{eq:coremassmin}, respectively. However, $\alpha$ is a parameter related to the properties of the host halo, as defined by eq.~\eqref{eq:def_alpha}. As shown in appendix~\ref{app:concentration}, when introducing the concentration parameter $c_{\rm vir} = r_{\rm vir}/r_{\rm s}$, the parameters $\rho_{\rm s}$ and $r_{\rm s}$ of the NFW density profile can be considered as variables determined by $c_{\rm vir}$ and the halo mass $M_{\rm h}$. Thus,by taking the limit of $c_{\rm vir} \to 0$, the maximum value of $r_{\rm c,max}(\alpha)$ and the minimum value of $M_{{\rm c, min}}(\alpha)$ for the given halo mass can be obtained as

\begin{align}
    r_{\rm c,max}^{\rm lim}&\equiv\lim_{c_{\rm vir}\to0}r_{\rm c,max}(\alpha)
    \nn
    &=3.80~[{\rm kpc}]~a^{-1/3}\left(\frac{\zeta(a)}{355}\right)^{-2/9}\left(\frac{\Omega_{\rm m,0}h^2}{0.126}\right)^{-2/9}\left(\frac{m_\psi}{8\times10^{-23}{\rm eV}}\right)^{-2/3}\left(\frac{M_{\rm h}}{10^9 M_\odot}\right)^{-1/9}\,,
    \label{eq:rcminlim}
    \\
    M_{\rm c,min}^{\rm lim}&\equiv\lim_{c_{\rm vir}\to0} M_{\rm c,min}(\alpha)
    \nn
    &=1.55\times 10^7~[M_\odot] ~a^{-2/3}\left(\frac{\zeta(a)}{355}\right)^{2/9} \left(\frac{\Omega_{\rm m,0}h^2}{0.126}\right)^{2/9}\left(\frac{m_\psi}{8\times10^{-23}{\rm eV}}\right)^{-4/3}\left(\frac{M_{\rm h}}{10^9 M_\odot}\right)^{1/9}\,,
    \label{eq:mcminlim}
\end{align}
where we have set the reference values of the cosmological parameters as $\Omega_{\rm m,0}=0.276,h=0.7$. These equations provide universal upper limits for $r_{\rm c}$ and lower limits for $M_{\rm c}$ that are independent of $\beta$ and $c_{\rm vir}$. In other words, given the FDM mass and halo mass, the core radius does not exceed $r^{\rm lim}_{\rm c,max}$, and the core mass does not fall below $M_{{\rm c, min}}^{\rm lim}$. Therefore, these limits are expected to be useful as criteria for determining the FDM mass from observational data. 
\section{Numerical analysis}
\label{sec:numerical_results}

In this section, we present the numerical solutions of the Schr\"{o}dinger-Poisson equations eq.~\eqref{eq:schrodinger2} and eq.~\eqref{eq:poisson2}. After describing the numerical method in section~\ref{sec:numerical_method}, the ground-state solution is numerically obtained for various set of parameters in section~\ref{sec:numerical_solution_of_soliton}, and the results are compared with the two limiting cases discussed in previous section.

\subsection{Method}
\label{sec:numerical_method}
Here, following ref.~\cite{Marsh:2015wka},  
we describe the numerical method to find the ground-state eigenfunction of the Schr\"{o}dinger-Poisson equations eq.~\eqref{eq:schrodinger2} and eq.~\eqref{eq:poisson2} with the five boundary conditions eq.~\eqref{eq:boundarycondition1}-eq.~\eqref{eq:boundarycondition5}. Since it is not practical to implement the boundary condition at infinity in a numerical scheme, we introduce a finite boundary at $x_{\rm max}$. This allows us to approximate eq.~\eqref{eq:schrodinger2} around the large $x_{\rm max}$ as
\begin{align}
    \tilde{u}''(x_{\rm max}) \approx -\calE \tilde{u}(x_{\rm max})\,.
\end{align}
As a result, $\tilde{u}$ is exponentially damped around $x_{\rm max}$ as
\begin{align}
    \tilde{u}(x_{\rm max})\propto e^{-\sqrt{-\calE}x_{\rm max}}\,.
\end{align}
Similarly, the right hand side of eq.~\eqref{eq:poisson2} becomes negligible at large $x_{\rm max}$, and around $x_{\rm max}$ the soliton self-gravity potential takes the form
\begin{align}
    \tilde{\Phi}(x_{\rm max})\propto \frac{1}{x_{\rm max}}\,.
\end{align}
We also set a small but nonzero inner boundary $x_{\rm min}$ to avoid an unexpected divergence at the center. To sum up,
We adopt the alternative boundary conditions instead of eq.~\eqref{eq:boundarycondition1}-eq.~\eqref{eq:boundarycondition5} as follows~\cite{Marsh:2015wka}:
\begin{align}
    \tilde{u}(x_{\rm min})&=\sqrt{\beta}\,,
    \label{eq:bdrcon1}\\
    \tilde{u}'(x_{\rm min})&=0\,,
    \label{eq:bdrcon2}\\
    \tilde{\Phi}'(x_{\rm min})&=0\,,
    \label{eq:bdrcon3}\\
    \tilde{u}'(x_{\rm max})&=-\sqrt{-\calE}\tilde{u}(x_{\rm max})\,,
    \label{eq:bdrcon4}\\
    \tilde{\Phi}'(x_{\rm max})&=-\frac{\tilde{\Phi}(x_{\rm max})}{x_{\rm max}}\,.
    \label{eq:bdrcon5}
\end{align}

Eigenvalue $\calE$ and eigenstates $\tilde{u}(x),\tilde{\Phi}(x)$ are obtained numerically by using the shooting method. First, with the initial guess of the values of $\tilde{\Phi}(x_{\rm min})$ and $\calE$, the dimensionless Schr\"{o}dinger-Poisson equation eq.~\eqref{eq:schrodinger2} and eq.~\eqref{eq:poisson2} are solved outward from $x_{\rm min}$ to $x_{\rm match}$, where we have set the matching point as $x_{\rm match}=x_{\rm max}/2$. Second, by assuming the value of $\tilde{u}(x_{\rm max})$ and $\tilde{\Phi}(x_{\rm max})$, the dimensionless Schr\"{o}dinger-Poisson equation eq.~\eqref{eq:schrodinger2} and eq.~\eqref{eq:poisson2} are solved inward from $x_{\rm max}$ to $x_{\rm match}$. The inward and outward solutions generally do not smoothly connect at $x_{\rm match}$. The Newton-Rapthon method is then used to find the appropriate values of $\tilde{\Phi}(x_{\rm min})$, $\tilde{\Phi}(x_{\rm max})$, $\tilde{u}(x_{\rm max})$, and $\calE$, which give $\tilde{u}(x)$, $\tilde{\Phi}(x)$ and their derivatives connected at $x_{\rm match}$. This process is repeated until the convergence is achieved. In the numerical calculation presented below, we set $x_{\rm min}$ to $10^{-5}$. The outer boundary $x_{\rm max}$ is chosen such that the square of the wavefunction satisfies $\tilde{u}(x_{\rm max})^2=10^{-2}$.

Note that the convergence of the Newton-Raphson method is not always efficient. It becomes extremely slow or to an undesirable excited state if the initial values of $(\tilde{\Phi}(x_{\rm min}), \tilde{\Phi}(x_{\rm max}), \tilde{u}(x_{\rm max}), \calE)$ are not sufficiently close to the actual values. Therefore, for given parameters $(\alpha, \beta)$, the computations are performed using an iterative method as follows.
\begin{enumerate}
    \item First, the eigenvalue and the eigenstate of the linearized Schr\"{o}dinger equation eq.~\eqref{eq:TS_Schrodinger} are obtained numerically for the parameters $(\alpha, \beta_1)$, where $\beta_1 \ll \beta_{\rm crit}(\alpha)$. In the limit where the soliton self-gravity is neglected (limit B), the Schr\"{o}dinger equation is reduced to a linear system, as we mentioned in section~\ref{subsubsec:Limit_B}. Discretizing $x$ and replacing the derivatives at each point with finite differences, the linearized Schr\"{o}dinger equation eq.~\eqref{eq:airyfunc} is represented in a matrix form~\cite{Taruya:2022zmt}. By diagonalizing this matrix, the eigenvalue $ \calE $ and the eigenfunction $ \tilde{u}(x) $ are determined simultaneously. Then, the dimensionless core radius $\bar{x}_c(\alpha)$ is obtained from eq.~\eqref{eq:def_core_radius}.
    \item Provided the parameters $(\alpha, \beta_1)$, the Schr\"{o}dinger-Poisson equations eq.~\eqref{eq:schrodinger2} and eq.~\eqref{eq:poisson2} are next solved by using the shooting method and the Newton-Raphson method. The initial condition for the eigenvalue $\calE$ is set to the value obtained in step 1. The initial condition for $\tilde{\Phi}(x_{\rm min})$ is obtained by integrating the Poisson equation with eq.~\eqref{eq:fitting_formula} to give
    \begin{align}
        \tilde{\Phi}(x_{\rm min})\approx -\frac{\beta x_{\rm c}^2}{14c}\,,
    \end{align} 
    with $x_{\rm c}=\bar{x}_c(\alpha)$ obtained in step 1. The initial conditions for both $\tilde{\Phi}(x_{\rm max})$ and $\tilde{u}(x_{\rm max})$ are set to 0.
    \item Choosing the value of $\beta_2$ slightly larger than $\beta_1$, the Schr\"odinger-Poisson equations eq.~\eqref{eq:schrodinger2} and eq.~\eqref{eq:poisson2} are solved for the new parameter set $(\alpha, \beta_2)$, again using the shooting method and the Newton-Raphson method, with the initial guess of $(\tilde{\Phi}(x_{\rm min}), \tilde{\Phi}(x_{\rm max}), \tilde{u}(x_{\rm max}), \calE)$ obtained in the previous step. 
    We gradually increase the value of $\beta$, and  
    repeat this step until we reach at the desired value of $\beta$.
\end{enumerate}

\begin{figure}[tbp]
\begin{minipage}[b]{0.5\linewidth}
    \centering
\includegraphics[width=\linewidth]{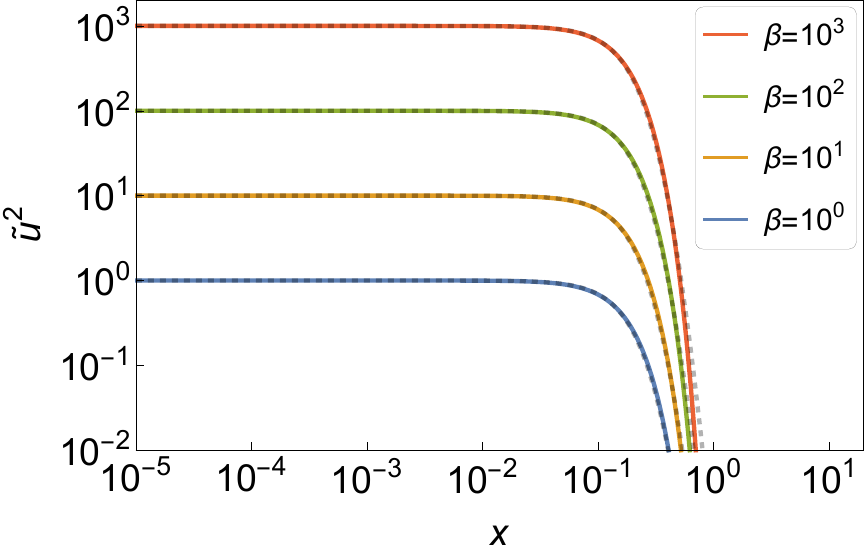}
\end{minipage}
\begin{minipage}[b]{0.5\linewidth}
    \centering
\includegraphics[width=\linewidth]{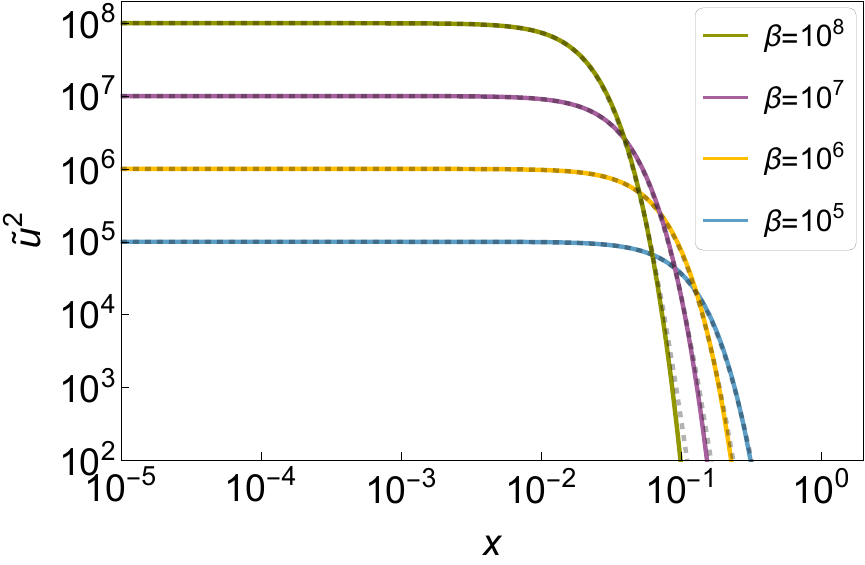}
    \end{minipage}
\caption{
    The dimensionless soliton density profile. The solid lines represent the numerical results with fixing $\alpha=10^4$. The colors represent the fixed values of $\beta$. The black dashed lines represent the fitting formula for the soliton density profile eq.~\eqref{eq:fitting_formula}. The left panel shows the result in the parameter region where the limit B is valid $(\beta\ll\beta_{\rm crit})$, while the right panel shows the results in the region where the limit A is valid $(\beta\gg\beta_{\rm crit})$.}
\label{fig:various_beta}
\end{figure}

\begin{figure}[tbp]
    \begin{minipage}[b]{0.5\linewidth}
    \centering
    \includegraphics[width=\linewidth]{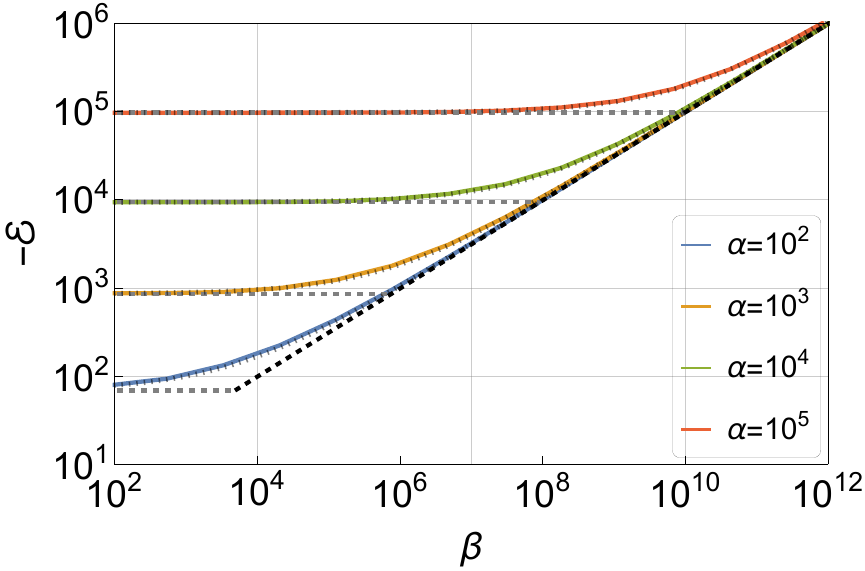}
    \end{minipage}
    \begin{minipage}[b]{0.5\linewidth}
    \centering
    \includegraphics[width=\linewidth]{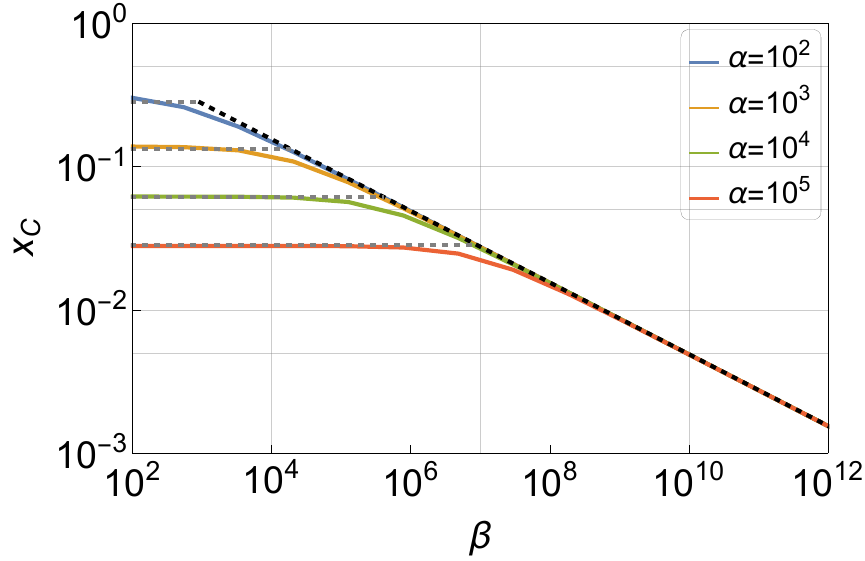}
    \end{minipage}
    \caption{
        The plots of the energy eigenvalue $-\calE$ (left) and the core radius $x_{\rm c}$ (right). The solid lines represent the numerical results, while the gray and black dashed lines represent the limit values given in the limit A (eq.~\eqref{eq:calEalpha} for $-\calE$ and eq.~\eqref{eq:TS_core_rad} for $x_{\rm c}$) and in the limit B (eq.~\eqref{eq:EnoSG} for $-\calE$ and eq.~\eqref{eq:coreradius} for $x_{\rm c}$), respectively. In the both panels, $\alpha={10^2~({\rm blue}), 10^3~({\rm yellow}), 10^4~({\rm green}), 10^5~({\rm red})}$ is fixed and $\beta$ is varied. The dotted lines in the left panel represent eq.~\eqref{eq:calEsum}.
        }
    \label{fig:calE}
\end{figure}

\begin{figure}[tbp]
    \begin{subfigure}[b]{0.5\linewidth}
    \centering
    \includegraphics[width=\linewidth]{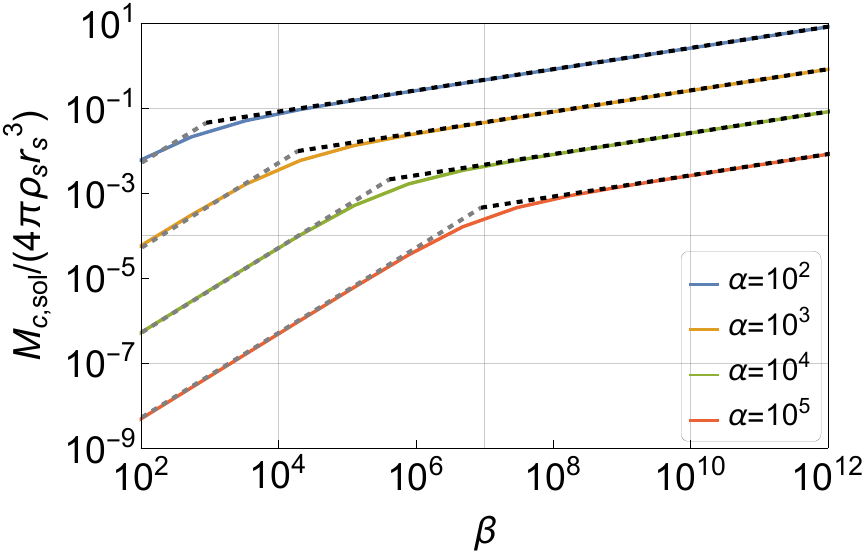}
    \caption{Core mass of the soliton core}
    \end{subfigure}
    \begin{subfigure}[b]{0.5\linewidth}
    \centering
    \includegraphics[width=\linewidth]{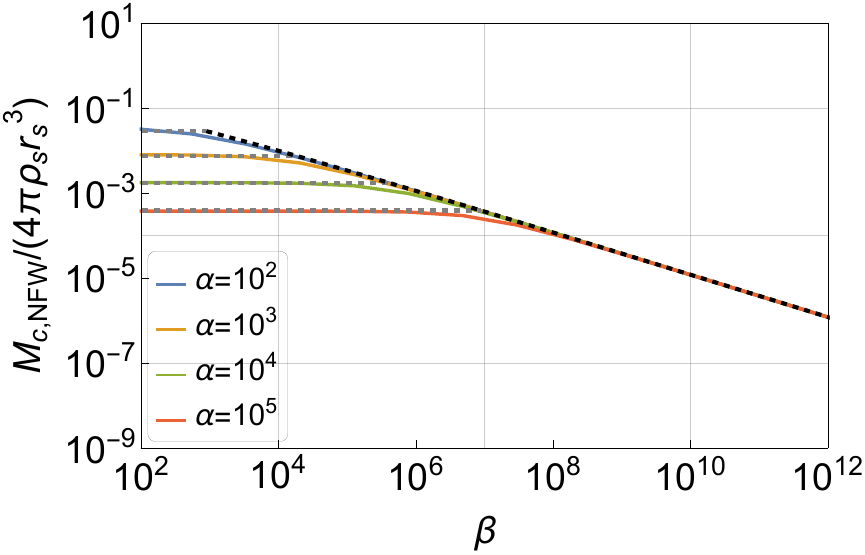}
    \caption{Core mass of the host halo}
    \end{subfigure}\\
    \begin{subfigure}[b]{\linewidth}
    \centering
    \includegraphics[width=0.5\linewidth]{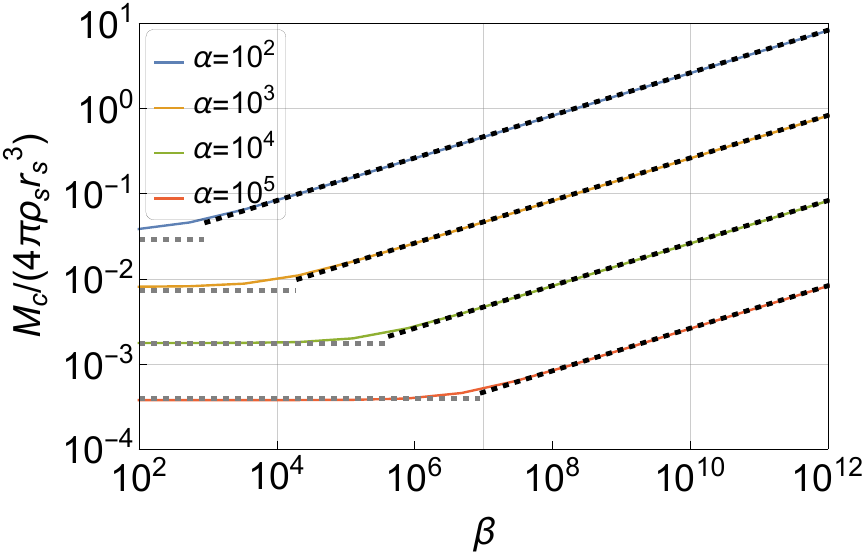}
    \caption{Core mass of the soliton core and the host halo}
    \end{subfigure}
    \caption{
        The dependence of the dimensionless core mass on $\alpha$ and $\beta$. The solid lines represent the numerical results, and their colors represent the value of $\alpha$, i.e., $\alpha=10^2$ (blue), $\alpha=10^3$ (orange), $\alpha=10^4$ (green), and $\alpha=10^5$ (red). The gray and black dashed lines represent the analytical results in the limit A and B, respectively. (a) shows the core mass of the soliton core, and black and gray dashed lines represent eq.~\eqref{eq:Mccase1}. (b) shows the core mass of the host halo, and black and gray dashed lines represent eq.~\eqref{eq:Mccase2}. (c) shows the total core mass. The gray and black dashed lines represent eq.~\eqref{eq:Mccase}, which correspond to the limit B and A, respectively.
        }
    \label{fig:Mcsolnfw}
\end{figure}

\subsection{Results}
\label{sec:numerical_solution_of_soliton}

With the numerical methods described above, the ground-state wavefunction $\tilde{u}(x)$, the self-gravitational potential $\Phi(x)$, and the eigenvalue $\calE$ are simultaneously and uniquely determined for given parameters $(\alpha,\beta)$. 
Figure~\ref{fig:various_beta} shows the examples of the dimensionless soliton density $\tilde{u}(x)^2$ obtained by the above numerical method. The fitting formula of the soliton density profile in eq.~\eqref{eq:fitting_formula}, depicted as gray dashed lines, fits well with the soliton profiles obtained from the numerical calculation, regardless of the value of the parameters $(\alpha,\beta)$. As illustrated in the left panel of figure~\ref{fig:various_beta}, which corresponds to the cases where the host halo potential is dominant, the core radii do not depend on $\beta$,  consistent with eq.~\eqref{eq:TS_core_rad}. In contrast, the right panel of figure~\ref{fig:various_beta} presents the results where the soliton self-gravity potential is dominant, and manifests the dependence of the core radius on $\beta$, again consistent with eq.~\eqref{eq:coreradius}.

Figure~\ref{fig:calE} presents the numerical values of the dimensionless ground-state eigenvalue $-\calE$ (left) and the core radius $x_{\rm c}$ (right), plotted as a function of $\beta$. Each line corresponds to a different value of $\alpha$. As illustrated in the left panel of figure~\ref{fig:calE}, in the region where $\beta \ll \beta_{\rm crit}\sim\alpha^{4/3}$, the eigenvalue $-\calE$ remains constant regardless of the value of $\beta$, and fits well with the approximate expression at eq.~\eqref{eq:calEalpha}. On the other hand, in the region where $\beta \gg \beta_{\rm crit}\sim \alpha^{4/3}$, it fits well with eq.~\eqref{eq:EnoSG} regardless of the value of $\alpha$. Combining them, we find that the energy eigenvalue of our model can be well reproduced by the sum of eq.~\eqref{eq:EnoSG} and eq.~\eqref{eq:calEalpha}, given by (see black dotted lines):
\begin{align}
    \calE\approx-\alpha+\left(\frac{9\pi}{16}\alpha\right)^{2/3}-0.979\,\beta^{1/2}\,,
    \label{eq:calEsum}
\end{align}
which still provides a reosonable approximation at the intermediate region $\beta\sim\beta_{\rm crit}$.

The right panel of figure~\ref{fig:calE} shows the dimensionless core radius $x_{\rm c}$ given as a function  of $\beta$ for representative values of $\alpha$. The solid lines represent the results of the numerical calculations, while the gray and black dashed lines represent the dimensionless core radius in the limit A  and B, given respectively at eq.~\eqref{eq:coreradius} and eq.~\eqref{eq:TS_core_rad}. The asymptotic behaviors of the numerical results match the analytical expressions in the limit A, eq.~\eqref{eq:coreradius} and B, eq.~\eqref{eq:TS_core_rad}.

In Figure~\ref{fig:Mcsolnfw}, we also show the core mass of soliton core $M_{{\rm c,sol}}$ and the core mass of the NFW density $M_{{\rm c,NFW}}$, which are defined by eq.~\eqref{eq:mcsol} and eq.~\eqref{eq:mcnfw}, respectively. Solid lines represent the results of numerical calculations. The gray and black dashed lines represent the asymptotic results in the limit B, eq.~\eqref{eq:Mccase1} and in the limit A, eq.~\eqref{eq:Mccase2}, respectively.

\section{Parameter reconstruction and implication from simulation data}
\label{sec:parameter_reconstruction}

In this section, as an implication of our model of the core-halo system discussed in the previous sections, we apply the model predictions to  the solitons found in the numerical simulations of refs.~\cite{May:2021wwp} and \cite{Chan:2021bja}, and interpret their structural properties in relation to the characteristics of the host halos. Specifically, we compare the characteristics of each soliton in numerical simulations with the predictions of the core-halo model to derive the key parameters $ (\alpha, \,\beta) $ and the concentration parameter of the host halo, $c_{\rm vir}$, which are not directly measured from simulations. In section~\ref{sec:transition_radius}, we begin by describing the method to reconstruct parameters $\alpha$, $\beta$, and $c_{\rm vir}$. Then, in section~\ref{subsec:results}, reconstructed results from the simulation data are presented, and statistical properties of the core-halo structure is discussed. In section~\ref{sec:FDM_mass_reconstruction}, we will briefly show that the FDM mass can be also reconstructed from the dataset in the simulation.

\begin{figure}[tbp]
    \begin{subfigure}[b]{\linewidth}
    \centering
    \includegraphics[width=0.5\linewidth]{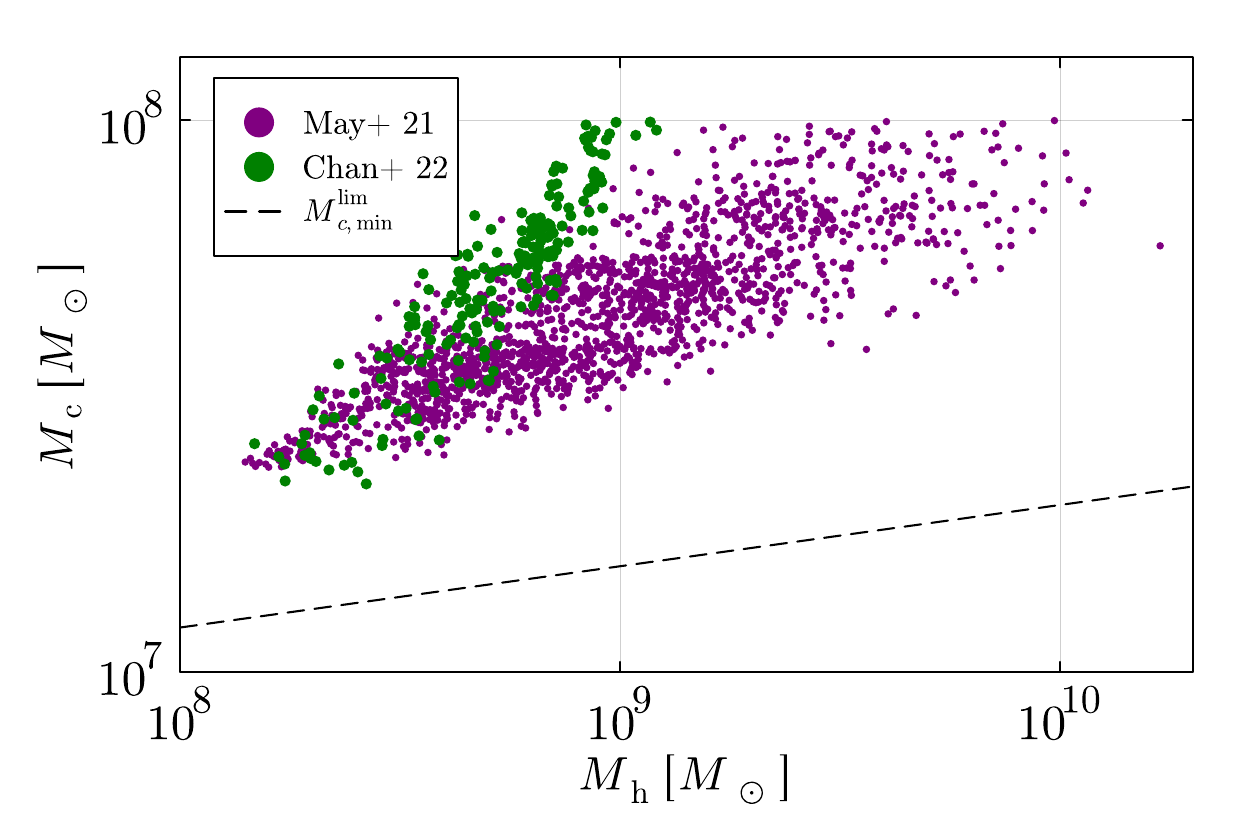}
    \caption{Core mass: $M_c$}
    \label{fig:reconstructed_cvir1}
    \end{subfigure}\\
    \begin{subfigure}[b]{0.5\linewidth}
    \centering
    \includegraphics[width=\linewidth]{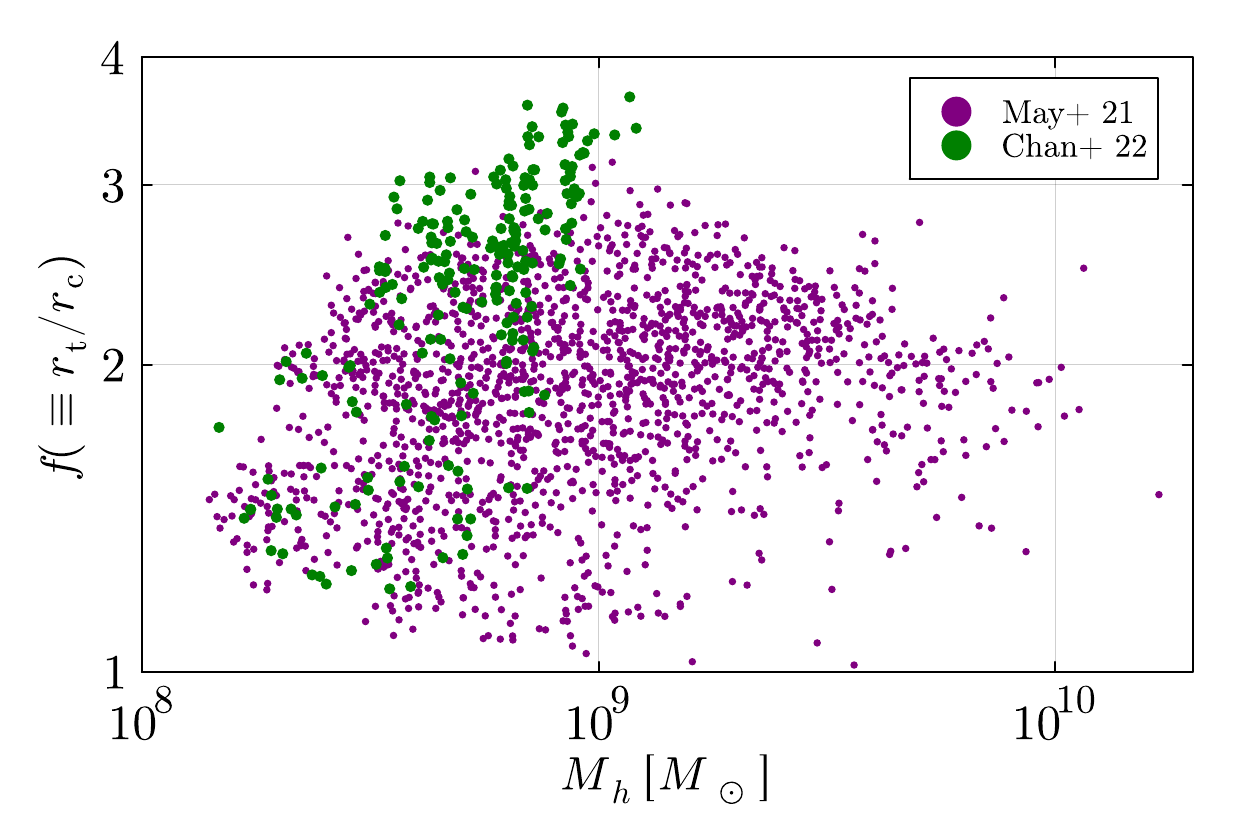}
    \caption{Fraction of the radius: $f\equiv r_{\rm t}/r_{\rm c}$}
    \label{fig:reconstructed_alpha1}
    \end{subfigure}
    \begin{subfigure}[b]{0.5\linewidth}
    \centering
    \includegraphics[width=\linewidth]{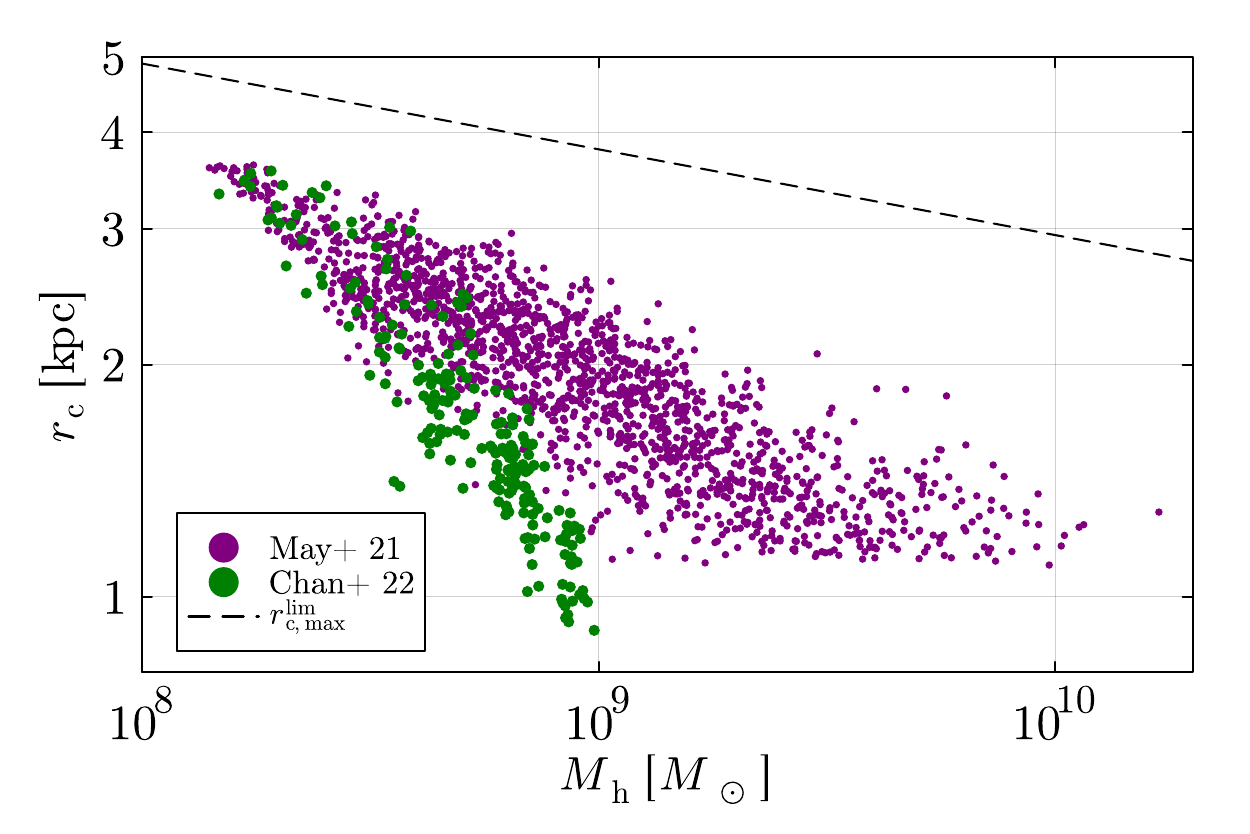}
    \caption{Core radius: $r_{\rm c}$}
    \label{fig:reconstructed_beta1}
    \end{subfigure}
    \caption{Input simulation data for the parameter reconstruction. The purple and green dots represent the result from the cosmological simulation~\cite{May:2021wwp} and soliton merger simulation~\cite{Chan:2021bja}, respectively.
    These figures are adopted by~\cite{Chan:2021bja}. In (a) and (c), the black dashed lines represent $M^{\rm lim}_{{\rm c,min}}$ and $r^{\rm lim}_{{\rm c, min}}$ respectively, which are introduced in section~\ref{app:coremass}.}
    \label{fig:Chuan_fig}
\end{figure}

\begin{figure}
    \centering
    \includegraphics[width=0.6\linewidth]{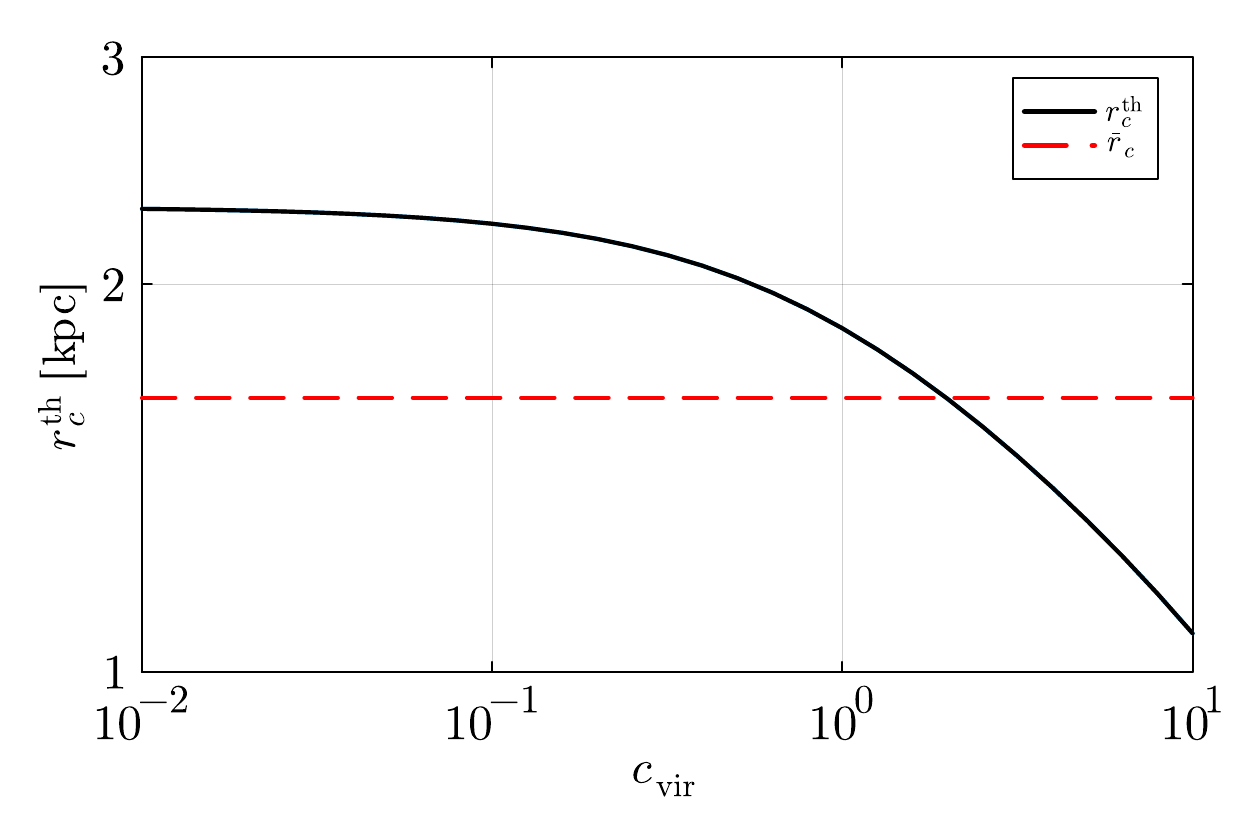}
    \caption{Black solid line represents the example of $r_{\rm c}^{\rm th}$ for a dataset in cosmological simulation. The red dashed line represents core radius $r_{\rm c}$ from the simulation.}
    \label{fig:rcth}
\end{figure}
\begin{figure}[tbp]
    \begin{subfigure}[b]{\linewidth}
    \centering
    \includegraphics[width=0.5\linewidth]{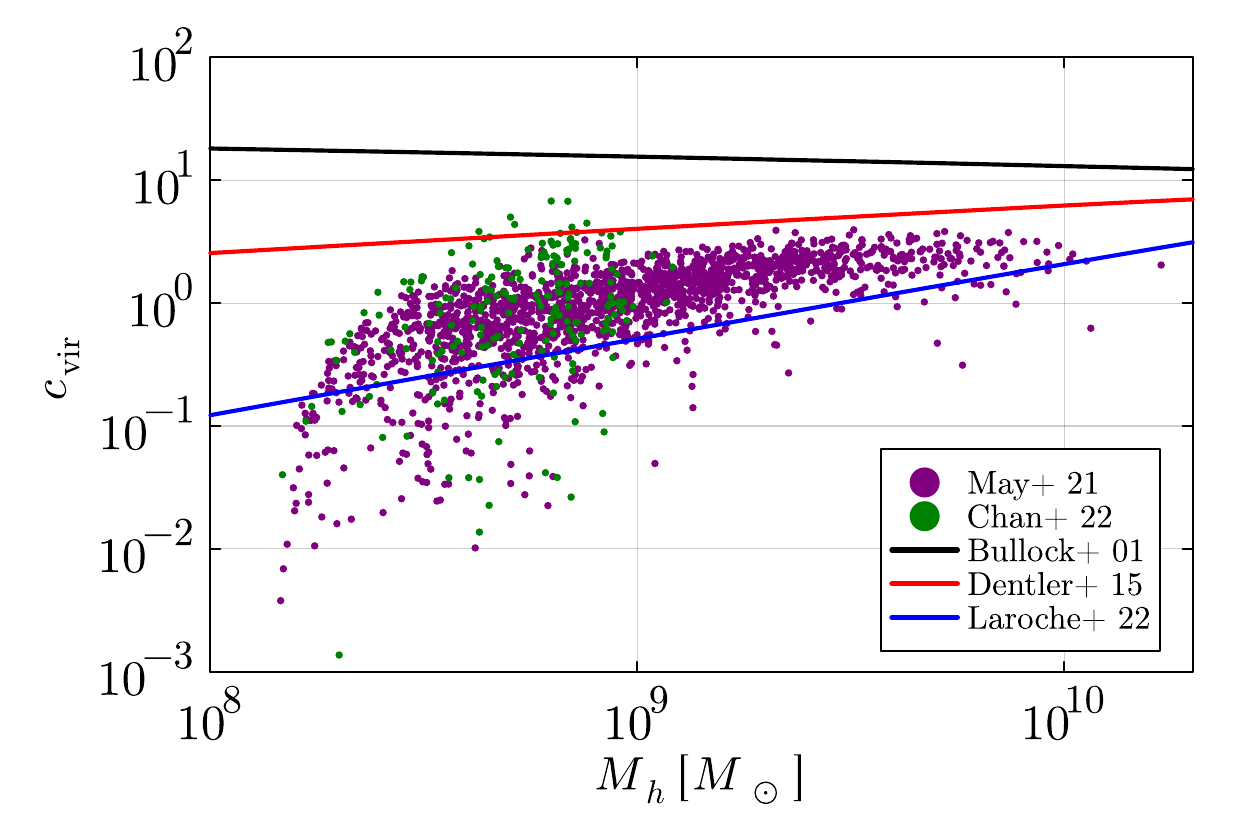}
    \caption{Reconstructed $c_{\rm vir}$.}
    \label{fig:reconstructed_cvir}
    \end{subfigure}\\
    \begin{subfigure}[b]{0.5\linewidth}
    \centering
    \includegraphics[width=\linewidth]{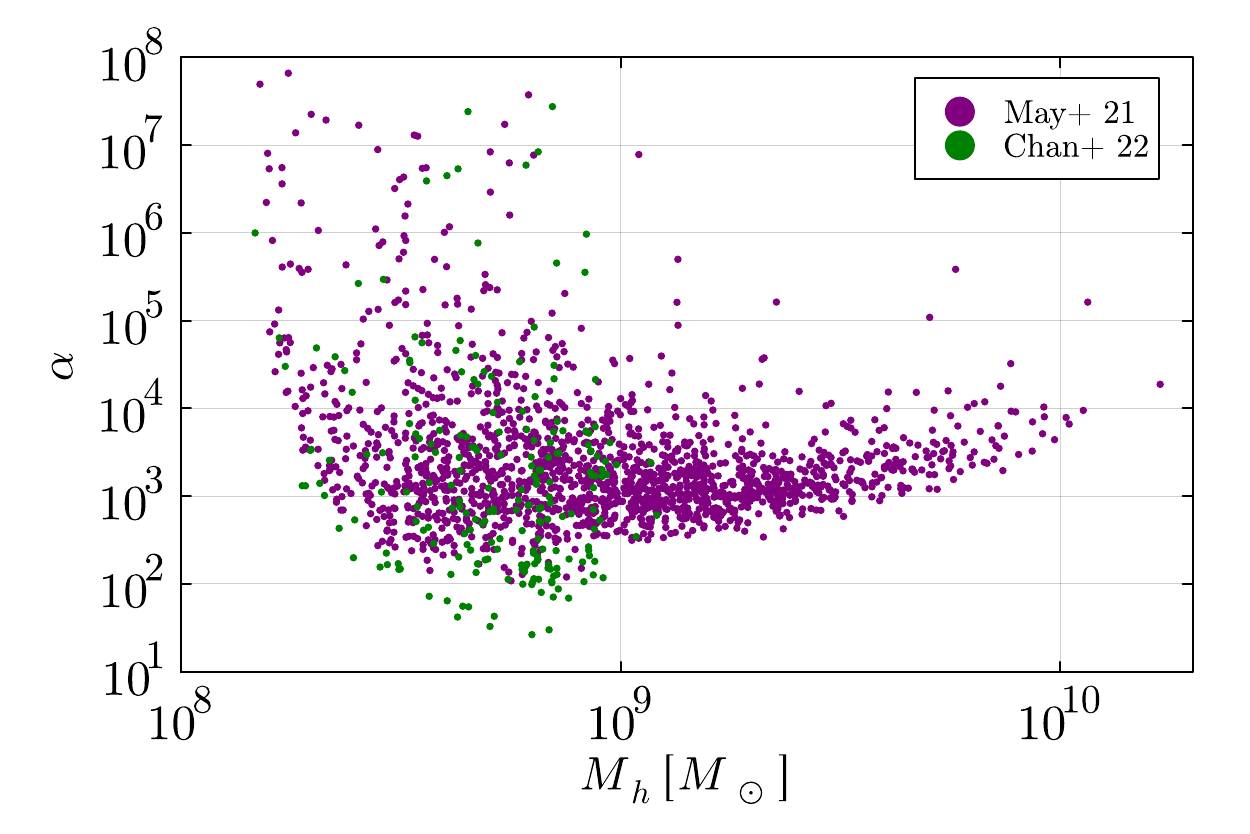}
    \caption{Reconstructed $\alpha$.}
    \label{fig:reconstructed_alpha}
    \end{subfigure}
    \begin{subfigure}[b]{0.5\linewidth}
    \centering
    \includegraphics[width=\linewidth]{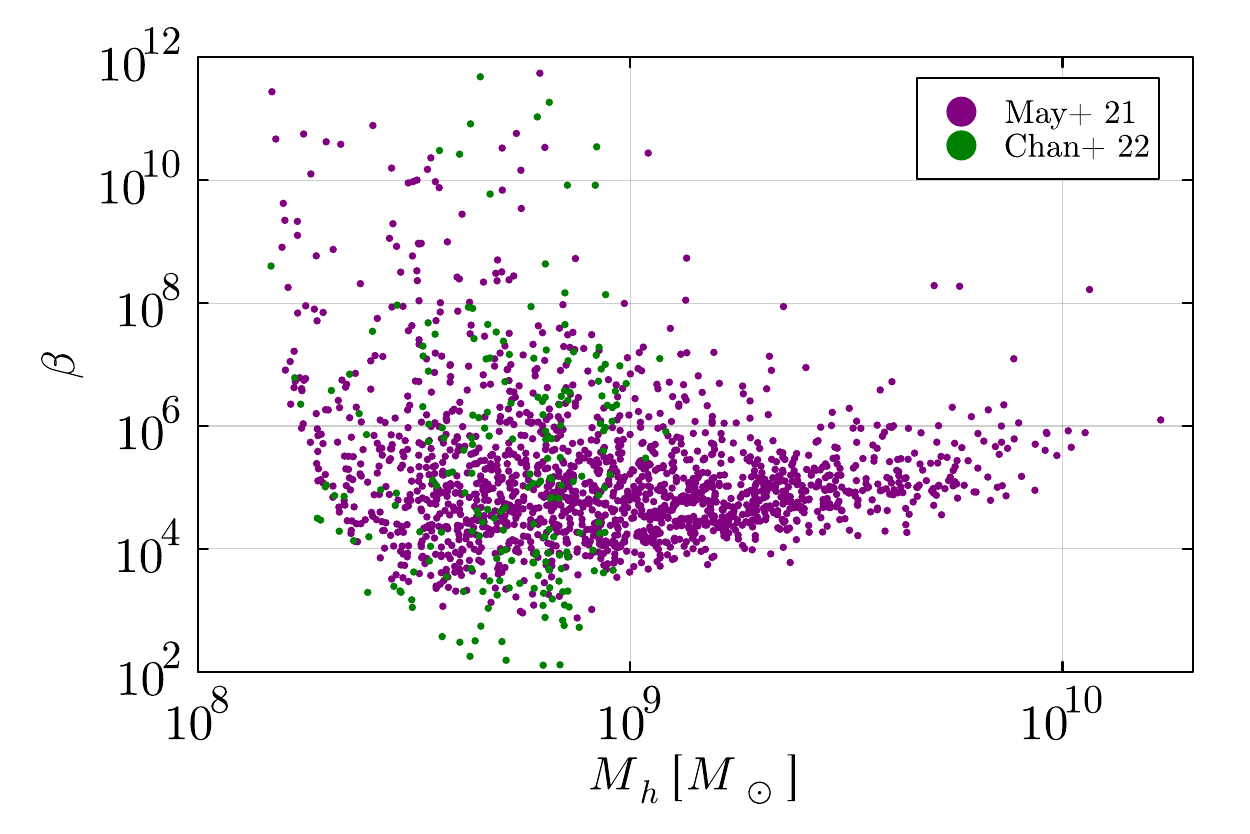}
    \caption{Reconstructed $\beta$.}
    \label{fig:reconstructed_beta}
    \end{subfigure}
    \caption{Reconstructed parameters $(c_{\rm vir}, \alpha, \beta)$ from the result of cosmological simulation (purple) and soliton merger simulation (green). (a) Relation between
    reconstructed concentration $c_{\rm vir}$ and the halo mass $M_{\rm h}$. Black and red lines show the concentration mass relation $c_{\rm vir}(M_{\rm h})$ for CDM halo~\cite{Bullock:1999he}, for FDM halo~\cite{Dentler:2021zij}. The blue line represent the another concentration mass relation for FDM halo~\cite{Laroche:2022pjm}.
    (b) Relation between reconstructed parameter $\alpha$ and the halo mass $M_{\rm h}$. (c) Relation between reconstructed parameter $\beta$ and the halo mass $M_{\rm h}$. }
    \label{fig:reconstructed_parames}
\end{figure}

\subsection{Formulation of the parameter reconstruction}
\label{sec:transition_radius}
Let us recall that for a given set of the parameters $(\alpha,\,\beta)$,
togerther with the host halo characteristics such as   
virial mass $M_{\rm h}$ and the concentration parameter $c_{\rm vir}$ (or scale radius $r_{\rm s}$), 
the model of the core-halo system can predict the soliton core characteristics such as core radius $r_{\rm c}$,  core density $\rho_{\rm c}$, and core mass $M_{\rm c}$ 
from the ground-state solution of 
the Schr\"odinger-Poisson equations, eq.~\eqref{eq:schrodinger2} and eq.~\eqref{eq:poisson2} in a fixed cosmological model. Combining the background halo profile, we can also obtain the total halo density profile, given as the sum of the soliton and the NFW profile. 
Conversely, we can read off the parameter $(\alpha,\,\beta)$ for each soliton from the halo and soliton characteristics, and one can investigate the actual properties of the core-halo system in numerical simulations.

Here, we use the data presented in ref.~\cite{Chan:2021bja}, consisting of the results obtained from soliton merger simulations and simulations started from the cosmological initial condition in ref.~\cite{May:2021wwp}. To be precise, we use the measured results of the soliton core radius $r_{\rm c}$, transition radius of the total profile $r_{\rm t}$, and the halo mass $M_{\rm h}$, as shown in figure~\ref{fig:Chuan_fig}. The transition radius $r_{\rm t}$ is defined as the scale where the density profile changes from a cored inner profile to the outer NFW profile. 
Specifically, in ref.~\cite{Chan:2021bja}, the transition radius $r_{\rm t}$ is determined by fitting the measured halo profile to the analytical fitting form consisting of the inner soliton profile $\rho_{\rm sol}^{\rm(fit)}$ at $r\leq r_{\rm t}$ (eq.~\eqref{eq:fitting_formula}) and the outer NFW profile $\rho_{\rm NFW}$ at $r_{\rm t}\geq r$ (eq.~\eqref{eq:rho_NFW})\footnote{In determining the transition radius $r_{\rm t}$, ref.~\cite{Chan:2021bja} assumes the relation at eq.~\eqref{eq:centralcoredensity}, which is strictly valid in the limit A case. However, as shown in appendix \ref{appendix:reconstructed_Mc}, our reconstructed $M_{\rm c}$  reproduce well the scaling relation with $r_{\rm c}$, which is derived from eq.~\eqref{eq:centralcoredensity} (see eq.~\eqref{eq:mcsol_unit}). In this respect, the results are self-consistent. }.

Given a set of parameters $(r_{\rm c},\,r_{\rm t}, \, M_{\rm h})$, the method to reconstruct the parameters $\alpha$ and $\beta$ is given as follows. 
Suppose that the cosmological model and the mass of FDM $m_\psi$ are a priori known\footnote{In the reconstructed procedure, the cosmological parameters are fixed as $a=1,\Omega_{m,0}=0.278, h=0.7$, and FDM mass is fixed as $m_\psi=8\times10^{-23}{\rm eV}$.}. 
eq.~\eqref{eq:def_alpha} implies that for a given halo mass, the parameter $\alpha$ is given as a single-variate function of the concentration parameter, $c_{\rm vir}$\footnote{The characteristic density $\rho_{\rm s}$ and scale radius $r_{\rm s}$ are rewritten with $M_{\rm h}$ and $c_{\rm vir}$ through eq.~\eqref{eq:rhos} and eq.~\eqref{eq:rs}, respectively. }. Also, the parameter $\beta$ is expressed as a function of $c_{\rm vir}$ by using the definition of the transition radius. From eq.~\eqref{eq:coredensity}, the relation $\rho_{\rm sol}^{\rm fit}(r_{\rm t})=\rho_{\rm NFW}(r_{\rm t})$ leads to
\begin{align}
     \beta= \alpha\,\frac{\bigl(1+c\,(r_{\rm t}/r_{\rm c})^2\bigr)^8}{(r_{\rm t}/r_{\rm s})\big(1+r_{\rm t}/r_{\rm s}\bigr)^2},
    \label{eq:eqfrac}   
\end{align}
with $c=0.091$. Substituting the information obtained from simulations into the right-hand side (i.e., $r_{\rm c}$, $r_{\rm t}$ and $M_{\rm h}$), the expression of $\beta$ above is shown to be dependent only on the concentration parameter $c_{\rm vir}$. Once the parameters $\alpha$ and $\beta$ are specified with $c_{\rm vir}$, our soliton-halo model predicts  the dimensionless core radius, $x_{\rm c}$, from which the core radius in physical units $r_{\rm c}$ is computed from
\begin{align}
     r_{\rm c}^{\rm th}(c_{\rm vir})= x_{\rm c}\bigl(\alpha(c_{\rm vir}),\,\beta(c_{\rm vir})\bigr)\,r_{\rm s}(c_{\rm vir}).
    \label{eq:rcth}   
\end{align}
Since the core radius $r_{\rm c}^{\rm th}$ is now entirely determined by the single parameter $c_{\rm vir}$, we can compare it with the measured value of $r_{\rm c}$ in simulations to derive a consistent estimate of $c_{\rm vir}$. This, in turn, allows us to determine the parameters $\alpha$ and $\beta$.

Figure~\ref{fig:rcth} demonstrates how one can obtain a consistent value of $c_{\rm vir}$ for a specific halo. The solid curve represents eq.~\eqref{eq:eqfrac} predicted based on the soliton-halo model, and the horizontal dashed line is the core radius measured in simulations. The intersection of these curves gives the solution we want, and from this consistent value of $c_{\rm vir}$ we can reconstruct $\alpha$ and $\beta$ for each halo from eqs.~\eqref{eq:def_alpha} and \eqref{eq:eqfrac}, respectively.

The reconstruction method described above is reduced to a single-parameter search, and as a bybroduct, we also obtain the concentration parameter for each halo, which is not actually presented in ref.~\cite{Chan:2021bja}. In this respect, one can investigate both the soliton core and halo properties in the FDM model. While this approach would enable an efficient parameter search, it requires solving the Schrödinger-Poisson equations multiple times to compute the dimensionless core radius $x_{\rm c}(\alpha,\,\beta)$. In appendix \ref{app:fitting}, to accelerate the parameter reconstruction, we build an accurate fitting formula for $x_{\rm c}$, and use it to evaluate the core radius in eq.~\eqref{eq:eqfrac}. 

\begin{figure}
    \centering
    \includegraphics[width=0.6\linewidth]{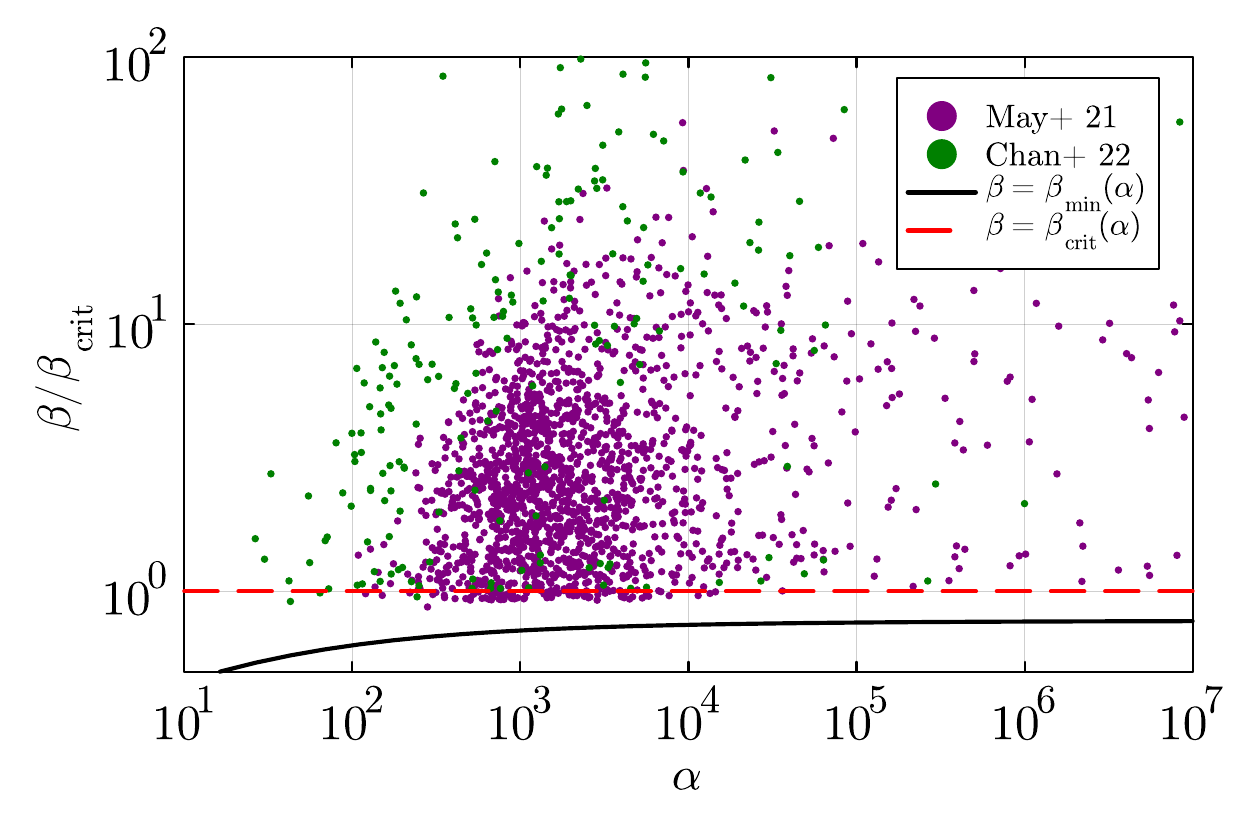}
    \caption{Relation between reconstructed two dimensionless parameters $\alpha,\beta$ from the result of cosmological simulation (purple) and soliton merger simulation (green). The red dashed line shows $\beta_{\rm crit}(\alpha)$ defined by eq.~\eqref{eq:goodapprox}. On the other hand, the black line represents $\beta_{\rm min}(\alpha)$.}
    \label{fig:reconstructed_alphabeta}
\end{figure}

\subsection{Results}
\label{subsec:results}

\subsubsection{Results of the parameter reconstruction}

Let us first look at the reconstructed parameters $(c_{\rm vir}, \alpha, \beta)$ given as a function of host halo mass, $M_{\rm h}$. In figure~\ref{fig:reconstructed_parames}, the purple and green dots represent the reconstruction results from the cosmological simulation and soliton merger simulation, respectively. In the data set we used, the soliton merger simulation tends to have a larger scatter than the cosmological simulation. 

In  figure~\ref{fig:reconstructed_cvir}, all of the reconstructed concentration parameter are found to be smaller than the one expected from CDM halos~\cite{Bullock:1999he}, depicted as black solid line. This is anticipated, to some extent, from the suppressed growth of small-scale structure with a large de Broglie wavelength. As a result, most of the reconstructed values lie between the two model prediction for FDM halos ~\cite{Dentler:2021zij} and~\cite{Laroche:2022pjm}, which are constructed by multiplying the suppression factor by the concentration-mass relation for CDM halos in ref.~\cite{Bullock:1999he}. The results indicate that the outer density profile of host halos is mostly dominated by a single power-law with the slope of $-1$.

On the other hand,  figure~\ref{fig:reconstructed_alpha} and \ref{fig:reconstructed_beta} respectively show the halo mass dependence of the reconstructed parameters $\alpha$ and $\beta$. We see a large scatter in both of the parameters. Especially for host halos of $M_{\rm h}\lesssim10^9\,$M$_\odot$, the reconstructed results of $\alpha$ and $\beta$ can reach a rather large value, $\sim10^8$ and $\sim10^{12}$, respectively, although they are not abundant but rather rare.

Figure~\ref{fig:reconstructed_alphabeta} shows the relationship between the reconstructed parameters $\alpha$ and $\beta$, with $\beta$ normalized by the critical value, $\beta_{\rm crit}$, defined at eq.~\eqref{eq:betacrit}. The black line represents the $\beta_{\rm min}(\alpha)$, which represents the minimum $\beta$ necessary for the existence of the transition radius for each $\alpha$ \footnote{The $\beta_{\rm min}$ is defined through eq.~\eqref{eq:eqfrac}, 
which is recast as
\begin{align}
    \frac{\beta}{\alpha} = \frac{(1 + c(x_t / x_c(\alpha,\beta))^2)^8}{x_t(1 + x_t)^2}\,,
\end{align}
where $x_{\rm t}\equiv r_t/r_c$.
For a given set of parameters $\alpha$ and $\beta$, this equation can be viewed as the transcendental equation for $x_t$. While this in general gives multiple solutions, the solution for $x_t$ becomes single-valued for specific parameters $\alpha$ and $\beta$, the latter of which is then defined as $\beta_{\rm min}$.
}. 
The reconstructed results for $\beta$ satisfy $\beta > \beta_{\rm min}(\alpha)$, as all simulated halos have a transition radius. Overall, there is no strong correlation between $\alpha$ and $\beta$, but most cases found in numerical simulations fall within the range $10^2 \lesssim \alpha \lesssim 10^4$, with the parameter $\beta$ ranging from 1 to $10^2$. This suggests that the self-gravity of the soliton core plays a crucial role in forming the cored density profile of the entire halo system, but the external contribution from the host halo to the soliton core cannot be completely ignored. In other words, the simulated core-halo system lies in between the limit A and limit B, rather than at either extreme.

    \begin{figure}
        \centering
        \includegraphics[width=\linewidth]{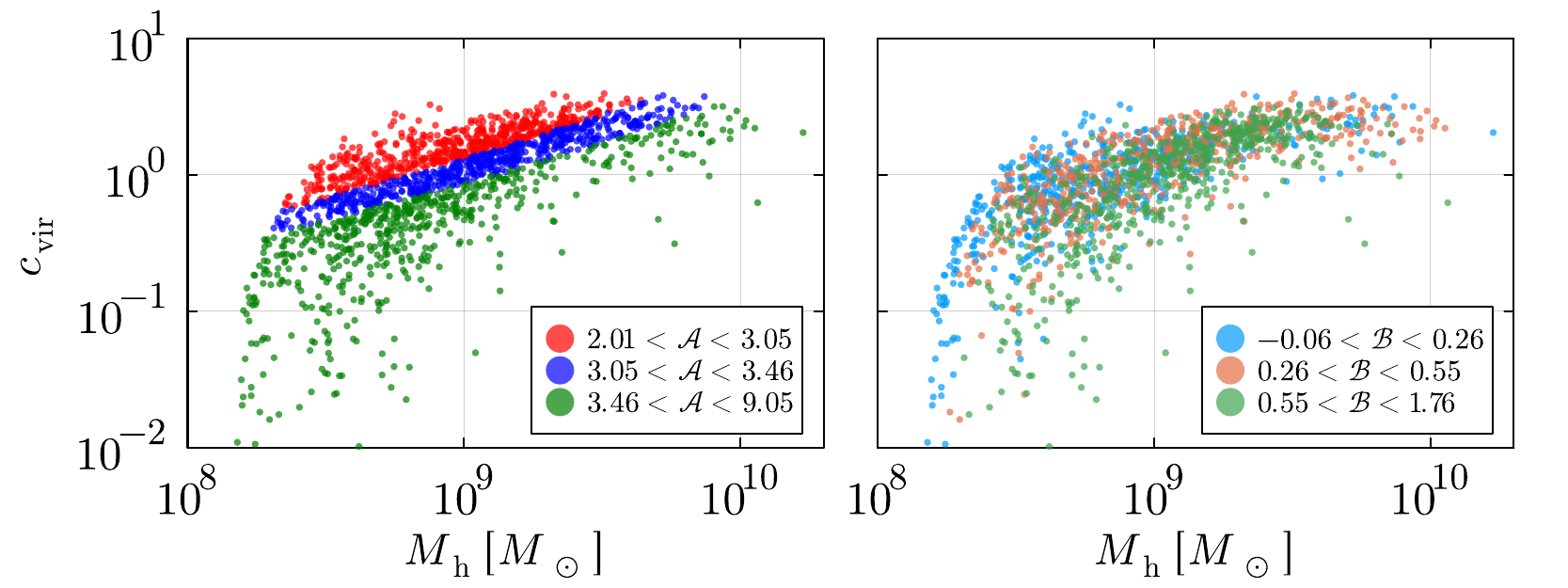}
        \caption{
        Relation between concentration parameter $c_\mathrm{vir}$ and halo mass $M_{\rm h}$ in the cosmological FDM halo simulation~\cite{May:2021wwp}. The colors represent the values of $\calA=\log_{10}\alpha$ (left) and $\calB=\log_{10}\beta/\beta_{\rm crit}$ (right).
        }
        \label{fig:physicalvariable_alpha_beta1}
    \end{figure}
    
    \begin{figure}
        \centering
        \includegraphics[width=\linewidth]{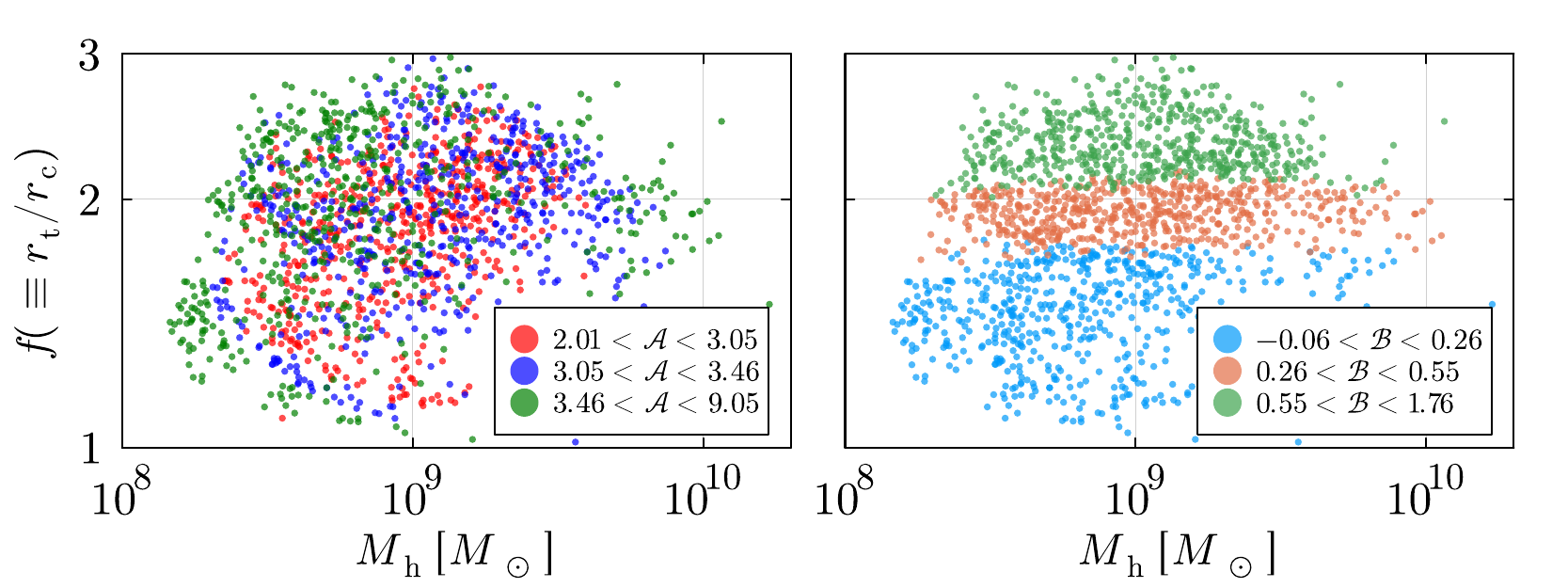}
        \caption{
        Relationship between the ratio of the transition radius to the core radius $f\equiv r_{\rm t}/r_{\rm c}$ and the halo mass $M_{\rm h}$ in the cosmological FDM halo simulation~\cite{May:2021wwp}. The colors represent the values of $\calA=\log_{10}\alpha$ (left) and $\calB=\log_{10}\beta/\beta_{\rm crit}$ (right).
        }
        \label{fig:physicalvariable_alpha_beta2}
    \end{figure}


\begin{figure}
    \centering
    \includegraphics[width=0.6\linewidth]{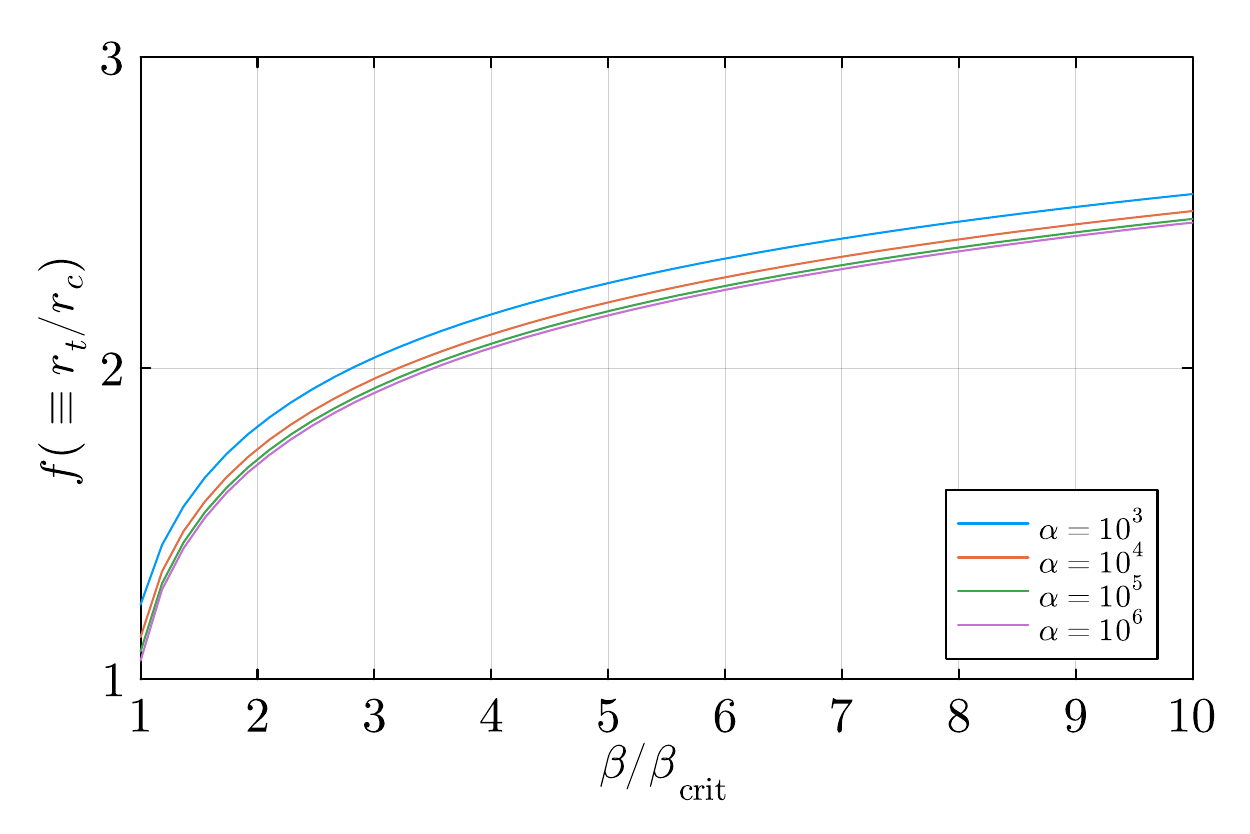}
    \caption{Dependence of the ratio of the transition radius to the core radius $f=r_{\rm t}/r_{\rm c}$ on $\beta/\beta_{\rm crit}$ and $\alpha$. Colors represent values of $\alpha$, i.e., $\alpha=10^3$ (blue), $10^4$ (orange), $10^5$ (green), $10^6$ (pink).}
    \label{fig:f_vs_beta}
\end{figure}
\subsubsection{Scatters in concentration parameter and transition radius}

In what follows, we focus only on the reconstructed results from the cosmological simulation since it contains a sufficient amount of data to analyze.

Figures~\ref{fig:physicalvariable_alpha_beta1} and \ref{fig:physicalvariable_alpha_beta2} show the relation between $c_{\rm vir}, f$ and the host halo mass $M_{h}$. The colors of each point represent the value of the reconstructed dimensionless parameters $\calA\equiv \log_{10}\alpha$ and $\mathcal{B}\equiv \log_{10}(\beta/\beta_{\rm crit})$, where $\calB$ indicates the magnitude of the self-gravity of the soliton core relative to that of the host halo. The reconstructed values of $\calA$ and $\calB$ are divided into three bins so that each bin has the same number of data in each of the left and right panels.

In figure~\ref{fig:physicalvariable_alpha_beta1}, we see that the concentration-mass relation $c_{\rm vir}(M_{\rm h})$ is strongly correlated with $\calA$, while no clear trend  is observed with $\calB$. This is understood explicitly from eq.~\eqref{eq:def_alpha}, eq.~\eqref{eq:rhos}, and eq.~\eqref{eq:rs}. We obtain 
\begin{align}
    \alpha \propto \left[c_{\rm vir}\log(1+c_{\rm vir})-\frac{c_{\rm vir}^2}{1+c_{\rm vir}}\right]^{-1} M_{\rm h}^{4/3}\,.
\end{align}
That is, a high-concentration halo tends to have a smaller value of $\alpha$. Also, low-mass halos tend to have a small  concentration parameter, consistent with figure~\ref{fig:physicalvariable_alpha_beta1}.

On the other hand, figure~\ref{fig:physicalvariable_alpha_beta2} shows the relationship between $f$ and $M_{\rm h}$. Here, each point is classified based on the dimensionless parameters $\calA$ (left) and $\calB$ (right), which are divided into three bins and represented by different colors. These plots show that $f(M_{h})$ is strongly correlated with $\calB$, but has no clear correlation with $\calA$. These trends can be understood qualitatively from figure~\ref{fig:f_vs_beta}, where we see clearly the monotonic increase of $f$ with $\beta/\beta_{\rm crit}$, while its increase is hardly changed with $\calA$.

\subsubsection{Scatter in core-halo relation}

\begin{figure}
        \centering
        \includegraphics[width=\linewidth]{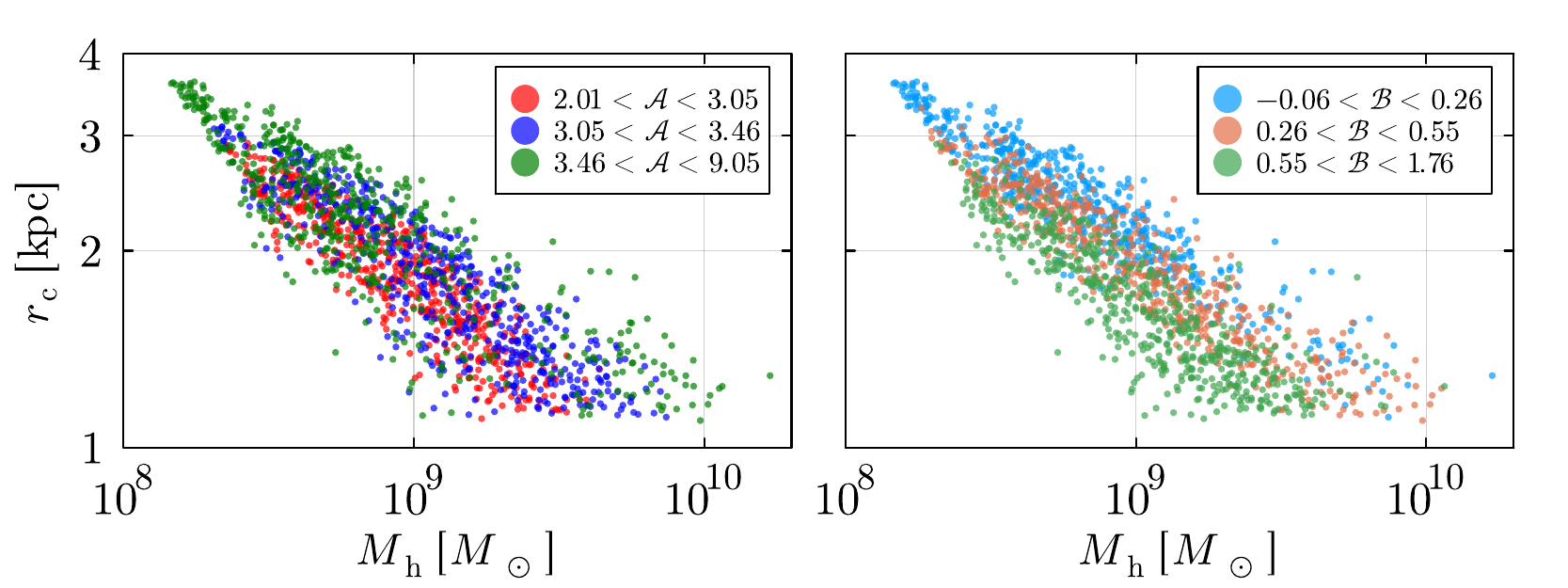}
        \caption{
        Relation between core radius $r_{\rm c}$ and halo mass $M_{\rm h}$ in the cosmological FDM halo simulation~\cite{May:2021wwp}. The colors represent the values of $\calA=\log_{10}\alpha$ (left) and $\calB=\log_{10}\beta/\beta_{\rm crit}$ (right).
        }
        \label{fig:physicalvariable_alpha_beta3}
\end{figure}

\begin{figure}
\begin{subfigure}{\linewidth}
    \centering
    \includegraphics[width=\linewidth]{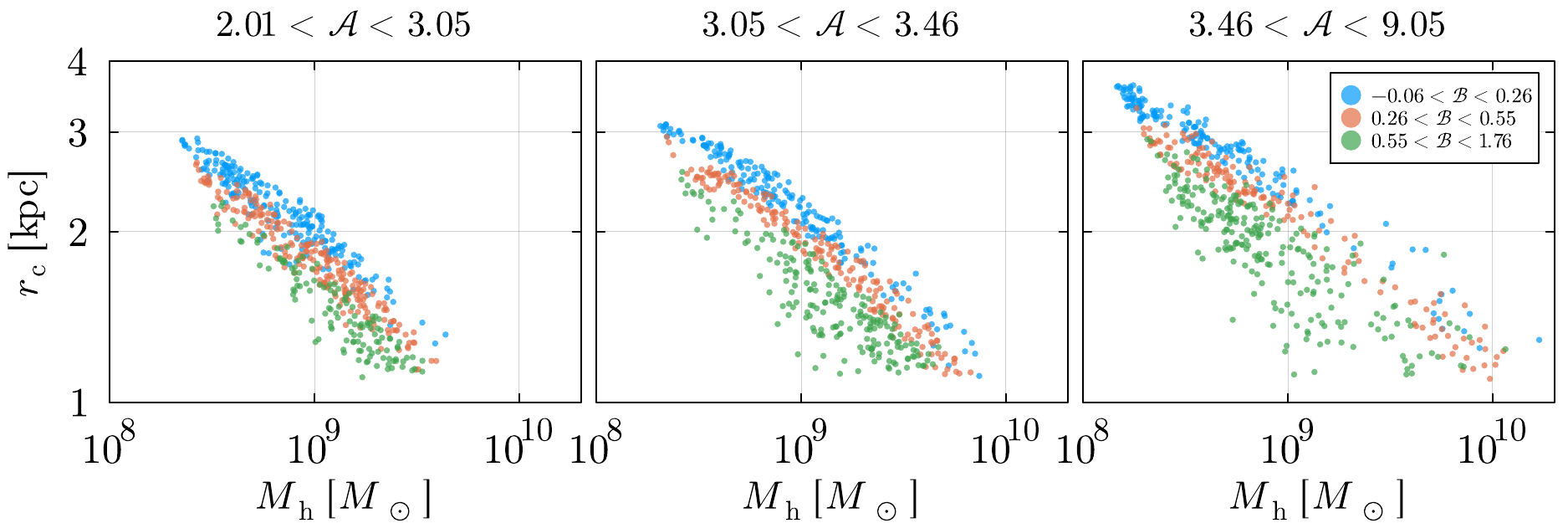}
    \phantomsubcaption
    \label{fig:rcalphabeta3}
\end{subfigure}\\
\begin{subfigure}{\linewidth}
    \centering
    \includegraphics[width=\linewidth]{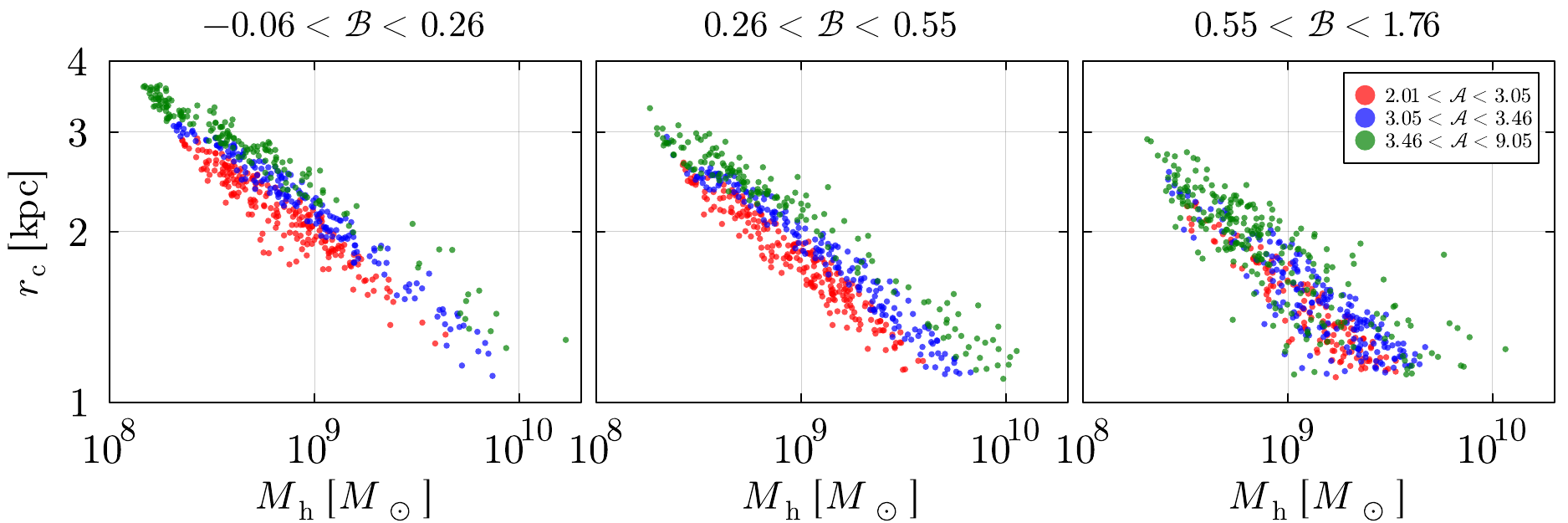}
    \phantomsubcaption
    \label{fig:rcbetaalpha3}
\end{subfigure}
\caption{
    Relationship between $r_{\rm c}$ and $M_{\rm h}$ in the cosmological simulation~\cite{May:2021wwp}. The data binned by $\calA\equiv\log_{10}(\alpha)$ (or $\calB$) was further binned by $\calB\equiv\log_{10}(\beta/\beta_{\rm crit})$ (or $\calA$). The colors represent the value of $\calA$ (or $\mathcal{B}$).
}
\label{fig:rcbetaalpha}
\end{figure}

Figure~\ref{fig:physicalvariable_alpha_beta3} shows the relationship between $r_{\rm c}$ and  $M_{\rm h}$ classified with $\calA$ (left) and $\calB$ (right) by three different colors. The left and right panels suggest a trend where the core radius tends to decrease as $\calA$ decreases or $\calB$ increases, although these trends appear subtle. To clarify the trend in more comprehensive way, we further classify the relation between $r_{\rm c}$ and $M_{\rm h}$ both with $\calA$ and $\calB$, as shown in figure~\ref{fig:rcbetaalpha}. In other words, the scatter in the core-halo relation is closely related to the scatter in both $\calA$ and $\calB$. 

As discussed in the previous section, given that $c_{\rm vir}$ and $f$ correlate with $\calA$ and $\calB$, respectively, figure~\ref{fig:rcbetaalpha} suggests that the scatter in the core-halo relation originates from the scatter in both $c_{\rm vir}$ and $f$. In other words, this implies that the scatter in the core-halo relation is due to the unique characteristics of both the halo and the soliton.

Previous research has examined whether the scatter in the core-halo relationship can be explained by the scatter in $c_{\rm vir}$ in N-body simulations of halos, which originates from the history of halo merging and mass accretion~\cite{Taruya:2022zmt,Kawai:2023okm}. Our results suggest that the scatter in the core-halo relation arises not only from the halo's merger and mass accretion history but also from the local formation and evolution history of the soliton at the halo center.

\subsection{FDM mass reconstruction}
\label{sec:FDM_mass_reconstruction}
We have so far investigated the properties of the parameters 
$\alpha$, $\beta$ and  $c_{\rm vir}$ from the dataset in the simulation results. In this subsection, as a possible application of the present core-halo model to  observational studies, we allow the FDM mass to be free, and demonstrate that by adding the core mass $M_{\rm c}$ as an additional input, the mass $m_\psi$ can be estimated in a self-consistent manner using a procedure almost identical to the one presented in section~\ref{sec:transition_radius}. 

By fixing the halo mass $M_{\rm h}$ and cosmological parameters, $\rho_{\rm s}$ and $ r_{\rm s} $ are determined solely by $ c_{\rm vir} $. Thus, when the FDM mass $ m_\psi $ is a free parameter, $ \alpha $ can be regarded as a function of $ c_{\rm vir} $ and $ m_\psi $, form its definition eq.~\eqref{eq:def_alpha}. Similarly, from eq.~\eqref{eq:eqfrac}, $ \beta $ also becomes a function of $ c_{\rm vir} $ and $ m_\psi $. Therefore, the core radius in eq.~\eqref{eq:rcth} can be extended as
\begin{align}
    r^{\rm th}_{\rm c}(c_{\rm vir},m_\psi)=x_{\rm c}(\alpha(c_{\rm vir},m_\psi),\beta(c_{\rm vir},m_\psi))r_{\rm s}(c_{\rm vir})\,,
    \label{eq:rcth_mass}
\end{align}
From eq.~\eqref{eq:mctotal}, the core mass is also calculated from $ c_{\rm vir} $ and $ m_\psi $ with given $(M_{\rm h},r_{\rm c},r_{\rm t})$ as
\begin{align}
    M_{\rm c}^{\rm th}(c_{\rm vir},m_\psi)&=4\pi b\frac{\beta(c_{\rm vir},m_\psi)}{\alpha(c_{\rm vir},m_\psi)}\rho_{\rm s}(c_{\rm vir}) r_{\rm c}^3\notag\\
    &+4\pi\rho_{\rm s}(c_{\rm vir}) r_{\rm s}(c_{\rm vir})^3 \left(-\frac{r_{\rm c}/r_{\rm s}}{1+r_{\rm c}/r_{\rm s}}+\log(1+r_{\rm c}/r_{\rm s})\right)\,,
    \label{eq:mcth_mass}
\end{align}
where we have used eqs.~\eqref{eq:coredensity} and \eqref{eq:mcsol_fit}. The conditions for consistency with the input dataset are $ r_{\rm c} = r_{\rm c}^{\rm th}(c_{\rm vir}, m_\psi) $ and $ M_{\rm c} = M_{\rm c}^{\rm th}(c_{\rm vir}, m_\psi) $. For a given set of $ (M_{\rm h}, r_{\rm t}, r_{\rm c}, M_{\rm c}) $, by solving these equations, $ c_{\rm vir} $ and $ m_\psi $ are reconstructed, and they uniquely determine the NFW parameters $ \rho_{\rm s}$ and $r_{\rm s} $ as well as the model parameters $ \alpha$ and $\beta$. Note that the input parameters used for the reconstructuction are those characterizing the total density profile of each halo. One thus expects that the method presented here can work well for the observations of dwarf spheroidal galaxies, where the information on the halo density profile is obtained through the Jeans analysis from the stellar velocity dispersion data.

\begin{figure}[tbp]
    \centering
    \includegraphics[width=0.6\linewidth]{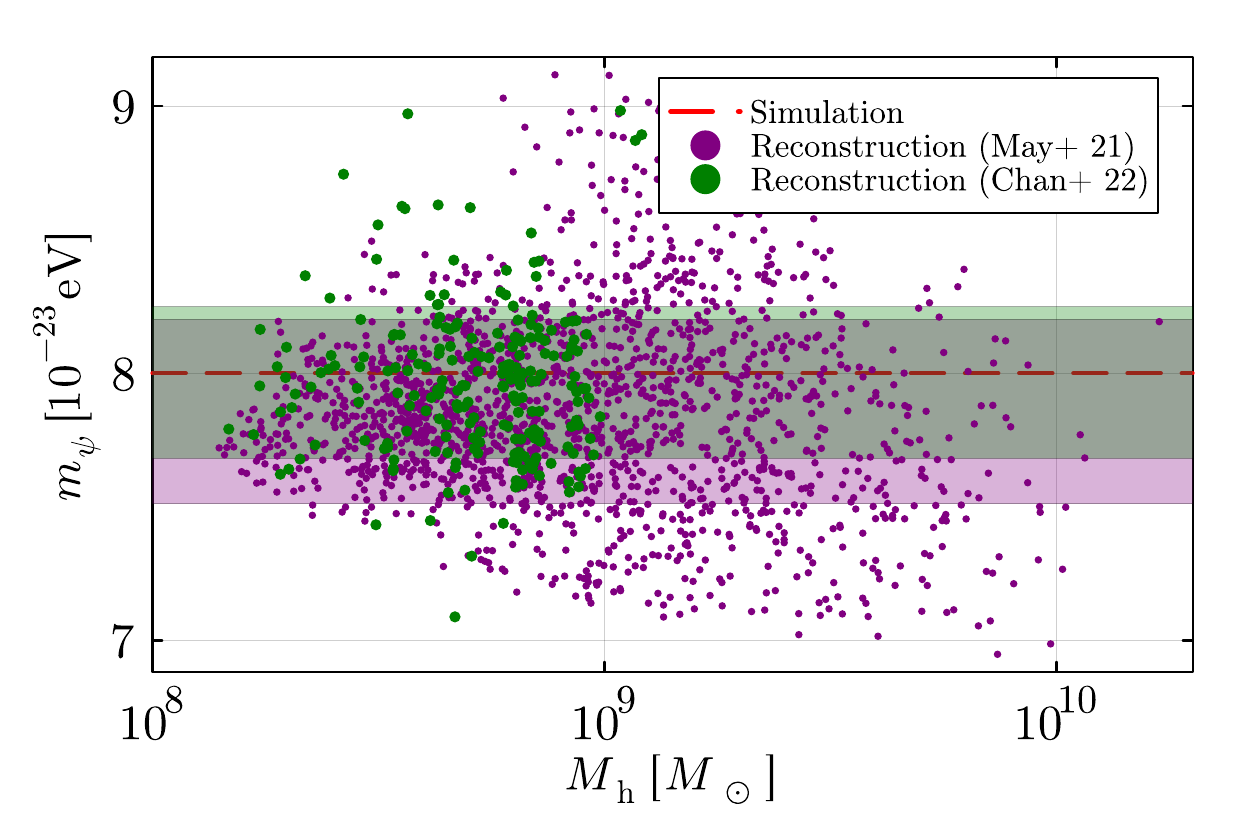}
    \caption{
     The results in the FDM mass reconstruction by the consistency of  eqs.~\eqref{eq:rcth_mass} and \eqref{eq:mcth_mass}.
     Red dahsed line represents the FDM mass set in these simulations, $m_\psi=8\times 10^{-23}~{\rm eV}$. We have used the set of data from the cosmological simulation \cite{May:2021wwp} and soliton merger simulation \cite{Chan:2021bja}, as an input data for the reconstruction procedure. 
    The purple and green stripes represent the $1\sigma$ region around the mean of the FDM mass data reconstructed from the cosmological and soliton merger simulation, respectively.
    }
    \label{fig:recm1}
\end{figure}

We examine the reconstruction method described above using a root finding algorism, where the initial conditions to search for solutions are set to the reconstructed value for $ c_{\rm vir} $ (see section~\ref{subsec:results}) and the fiducial value in  simulations for the FDM mass, i.e., $ m_\psi = 8 \times 10^{-23} \, {\rm eV} $. As a result, the reconstructed values of $ c_{\rm vir} $, $\alpha$ and  $\beta$ show no significant differences compared to those in figure~\ref{fig:reconstructed_parames}. The estimated FDM mass is shown in figure~\ref{fig:recm1}, where we see a good agreement with the fiducial value in the simulation to within about 10\% accuracy. 

Note that there are observational works that ignore the host halo contribution to the soliton core structure, and obtain the constraint on the FDM mass (e.g.~\cite{Safarzadeh:2019sre,Hayashi:2021xxu}). We can perform the reconstruction of the FDM mass in a similar setup, restricting our core-halo model to the Limit A. In this case, $ m_\psi $ can be reconstructed by substituting the simulation results $ (r_{\rm c}, M_{\rm c}) $ into eq.~\eqref{eq:mcsol_unit}. Comparing these results with those obtained in figure~\ref{fig:recm1}, we found that the reconstructed mass $m_\psi$ differs by only $\sim3\%$. This suggests that the scaling relation in eq.~\eqref{eq:mcsol_unit}, which ignores the contribution of the host halo, serves as a good approximation for the range of parameters we considered, despite the presence of solitons in the simulation results where $ \beta/\beta_{\rm crit} \sim \mathcal{O}(1)  $.

\section{Summary and discussion}
\label{sec:summary}
In this paper, we consider a model of the core-halo system in the FDM cosmology, described by the Sch\"{o}dinger-Poisson equation. In this system, the soliton core is supported by both a self-gravity of the soliton core itself and the host halo under the spherical and stationary ansatz. This system is characterized by the two dimensionless parameters $(\alpha,\beta)$, respectively given at eq.~\eqref{eq:def_alpha} and eq.~\eqref{eq:defbeta}. In regions where $\beta\gg\beta_{\rm crit}(\alpha)$, the limit that ignores the self-gravity of the host halo (limit A) is valid, while, in regions where $\beta\ll\beta_{\rm crit}(\alpha)$, the limit that ignores the soliton self-gravity (limit B) is valid. For a given $\alpha$, there exists a maximum value of the core radius and a minimum value of the core mass which correspond to the limit B. These values depend only on the FDM mass and the halo mass, and the concentration parameter through $\alpha$. As shown in section~\ref{app:coremass}, the limit $c_{\rm vir}\to0$ for a given FDM mass and halo mass gives the maximum value of $r_{\rm c, max}$ and the minimum value of $M_{{\rm c, min}}$, namely, they are universal bounds for a given FDM mass and halo mass, which are independent of neither $\beta$ nor $c_{\rm vir}$. These results are expected to be useful as a criterion for determining the FDM mass from observational data.

In addition, we have demonstrated that the dimensionless parameters $\alpha$ and $\beta$ can be reconstructed from the set of data in the results of the FDM halo simulations. As a result, most of the solitons in the FDM halo simulations are in the parameter range of $\beta/\beta_{\rm crit}\sim\mathcal{O}(1)-\mathcal{O}(10^2)$. This suggests that the soliton of the FDM halo is not only affected by the self-gravity of the soliton core itself, but also by that of the host halo. Further, the model parameters $\alpha$ and $\beta$ are strongly correlated with $c_\mathrm{vir}$ and $f$, respectively. Thus, it is suggested that the core-halo relation of the FDM halo depends on both the soliton characteristics, such as transition radius and core radius, and the host halo characteristics, such as the concentration parameter. We have also demonstrated that the FDM mass can be also reconstructed from $M_{\rm h},r_{\rm c},r_{\rm t},$ and $M_{\rm c}$. This method will be useful to get the constraint on FDM mass by applying the actual  data  in the future work.

As a final remark, our model can also be applied to investigate the baryonic effects on the soliton core by substituting the NFW density profile with the baryonic profile. In the context of self-interacting dark matter, it is known that baryonic effects contribute to the diversity of rotation curves in spiral galaxies \cite{Kamada:2016euw}. However, the influence of baryons on the halo structure in FDM remains poorly understood (but see ref.~\cite{Veltmaat:2019hou}), presenting an important direction for future research. 

\appendix

\section{Fitting formula of the dimensionless core radius}
\label{app:fitting}
\begin{figure}[tbp]
    \centering
    \includegraphics[width=0.6\linewidth]{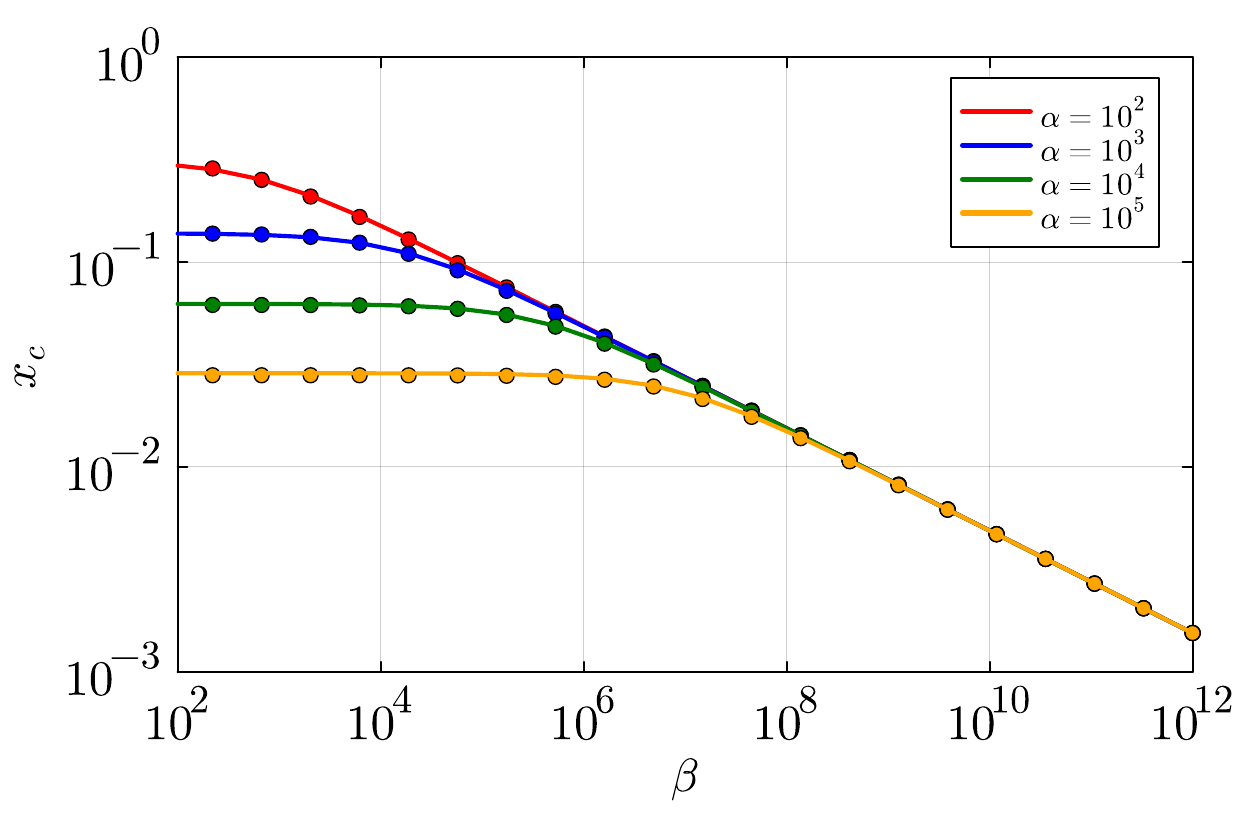}
    \caption{
    Solid lines represent the fitting formula of core radius given by eq.~\eqref{eq:xc_fitting}. Points represent the numerical values of $x_c$ by the shooting method shown in section~\ref{sec:numerical_method}.
    }
    \label{fig:xcbeta_fitting}
\end{figure}

In the process of reconstruction in section~\ref{sec:parameter_reconstruction}, $x_{\rm c}$ in eq.~\eqref{eq:rcth} can be calculated numerically using the shooting method, as demonstrated in the previous section. However, this numerical calculation have been found to be a bottleneck of the calculation time. Therefore, in the following section, we apply a fitting formula for the core radius as
\begin{align}
    x^{\rm fit}_c(\alpha,\beta) &= \calF(\alpha,\beta) (\beta^n+\beta_c^n)^{-\frac{1}{4n}}\,,
    \label{eq:xc_fitting}
    \\
    \calF(\alpha,\beta)&=q-\left(q-\beta_c^{1/4} \bar{x}(\alpha)\right)\left[1-\exp \left(-\frac{\beta}{\beta_c}\right)\right]\,.
\end{align}
The constant $n$ was set to $n=0.86$ to match the numerical calculations. The solid line in figure~\ref{fig:xcbeta_fitting} represents this fitting formula eq.~\eqref{eq:xc_fitting}, and the points represent the results of the numerical calculations using the shooting method.

\section{Concentration parameter}
\label{app:concentration}
We introduce the concentration parameter $c_{\rm vir}:=r_{\rm vir}/r_{s}$ to describe a NFW-type host halo.  Here, $r_{\rm vir}$ denotes the virial radius of the NFW density profile. The concentration parameter allows us to express $\rho_{\rm s}$ and $r_{\rm s}$ as
\begin{align}
    \rho_{\rm s}&=\frac{c_{\rm vir}^3}{3(\log(1+c_{\rm vir})-c_{\rm vir}/(1+c_{\rm vir}))}\zeta(a)\rho_{m,0}\,,
    \label{eq:rhos}
    \\
    r_{\rm s}&=\frac{1}{c_{\rm vir}}\left(\frac{3M_{\rm h}}{4\pi \zeta(a)\rho_{m,0}}\right)^{1/3}\,,
    \label{eq:rs}
\end{align}
where $M_{\rm h}$ is the NFW halo tail mass, and $\rho_{m,0}$ is the present matter density. Here, $\zeta$ is the redshift-dependent factor defined as
\begin{align}
    \zeta(a)=\frac{18\pi^2+82(\Omega_{\rm m}(a)-1)-39(\Omega_{\rm m}(a)-1)^2}{\Omega_{\rm m}(a)}\,.
\end{align}
Here, $\Omega_{\rm m}(a)$ is the matter density parameter at the scale factor $a$. 

A concentration-mass relation $c_{\rm vir}(M_{\rm h})$ depends on several factors such as the dark matter model, the initial power spectrum, cosmological parameters, redshift, and so on. While the concentration-mass relation for the CDM halo is numerically well studied by N-body simulations~(for example \cite{Bullock:1999he}), that for the FDM is relatively poorly understood, except for some analytical models~\cite{Dentler:2021zij,Laroche:2022pjm}. 

\section{Consistency check with the reconstructed core mass}
\label{appendix:reconstructed_Mc}

In section~\ref{sec:parameter_reconstruction} we show how to reconstruct the model parameters $\alpha$ and $\beta$ from given simulation results $(M_{\rm h}, r_{\rm t}, r_{\rm c})$. In our procedure, the core mass $M_c$ is not used as an input parameter because $M_c$ is related to $r_{\rm c}$ \cite{Chan:2021bja}.

\begin{figure}
    \centering
    \includegraphics[width=0.6\linewidth]{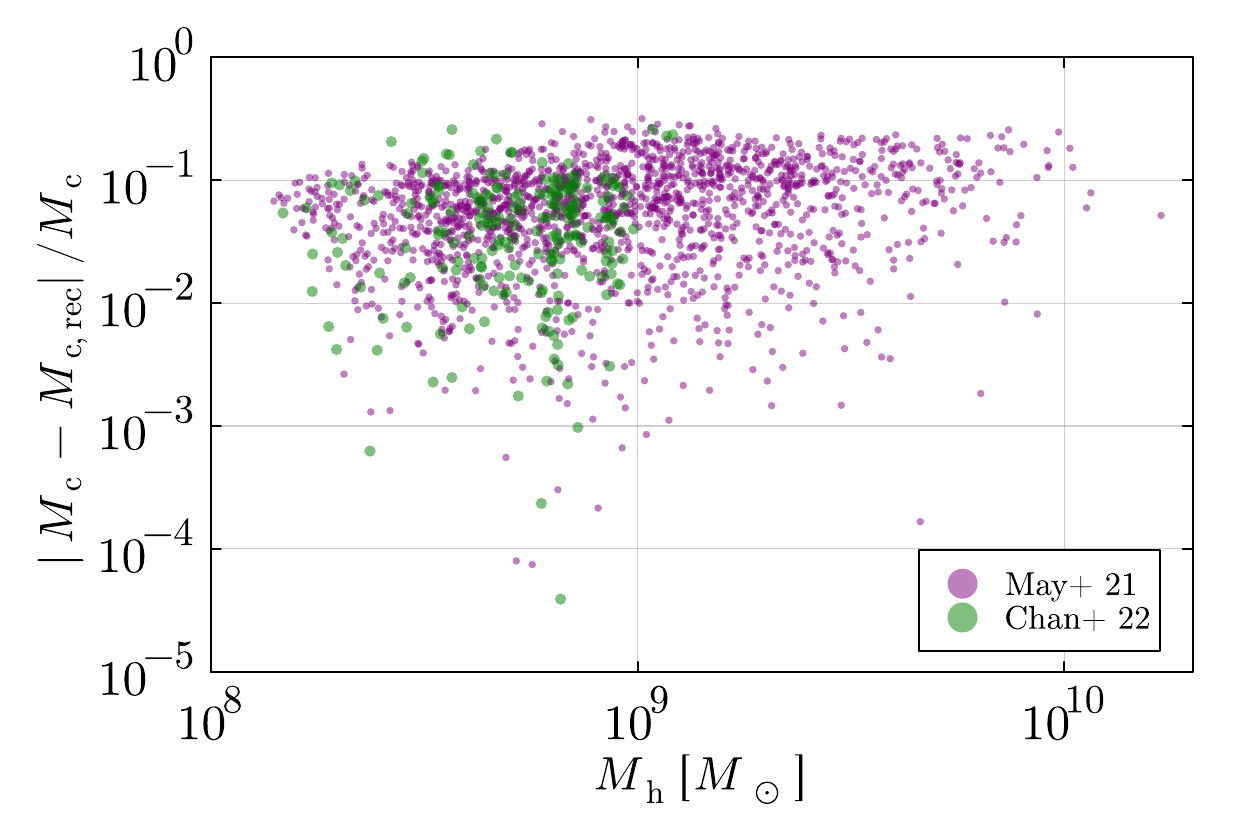}
    \caption{
    The relative difference between the core mass from the simulation results and the reconstructed core mass. We used data from cosmological (purple) and soliton merger (green) simulations~\cite{May:2021wwp, Chan:2021bja}. Here, $M_{\rm c}$ is an input core mass from the simulation result, and $M_{\rm c, rec}$ represents the reconstructed core mass.
    }
    \label{fig:relative_error}
\end{figure}

In our model, the core mass can be calculated from the reconstructed parameters $(\alpha,\beta,c_{\rm vir})$ and the input parameters $(M_{\rm h},r_{\rm t},r_{\rm c})$ from eqs.~\eqref{eq:mctotal}-\eqref{eq:fitting_formula}. Therefore, the reconstructed core mass can be used as a consistency check by comparing it with the core mass obtained from the simulation results. Figure~\ref{fig:relative_error} shows the relative difference between the core mass from the simulation results and the reconstructed core mass, showing that the core mass is correctly reconstructed up to $\sim30\%$ error.  

\begin{figure}
\begin{minipage}[b]{0.5\linewidth}
    \centering
    \includegraphics[width=\linewidth]{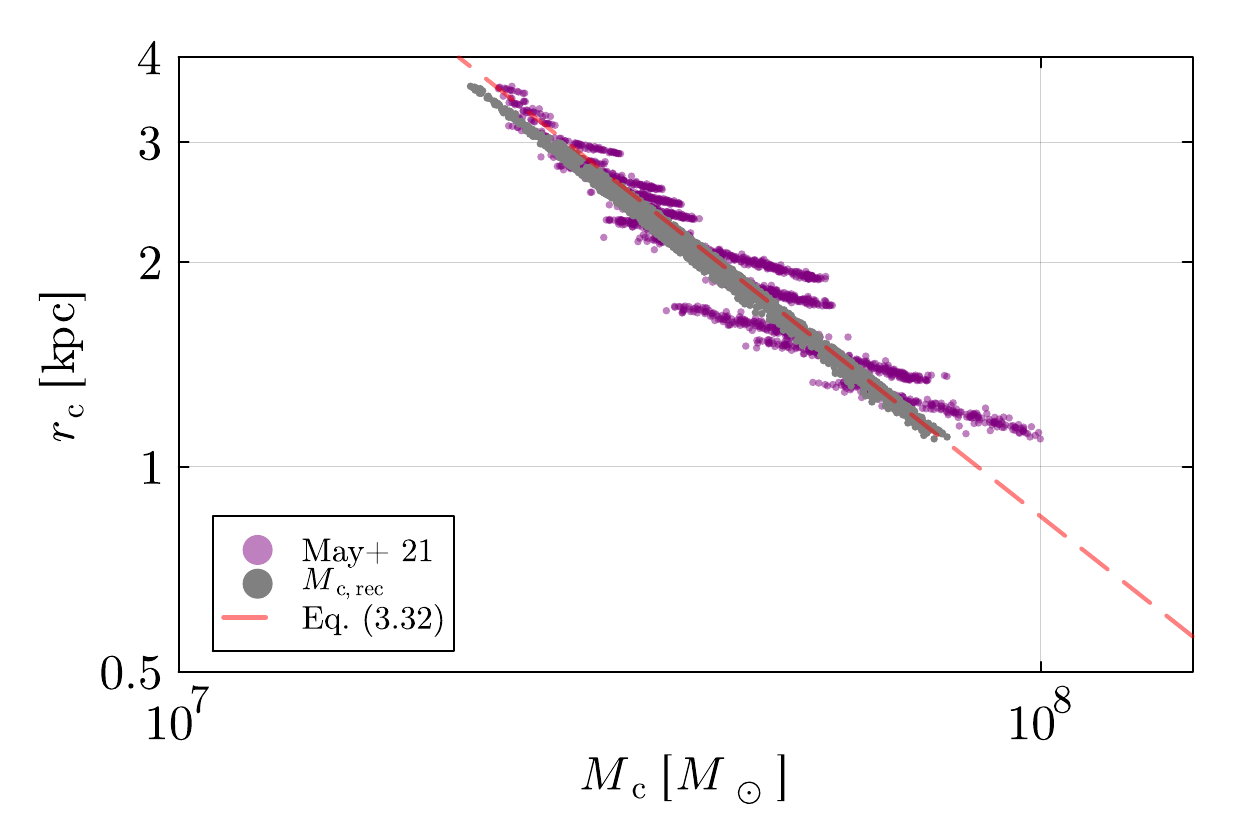}
\end{minipage}
\begin{minipage}[b]{0.5\linewidth}
    \centering
    \includegraphics[width=\linewidth]{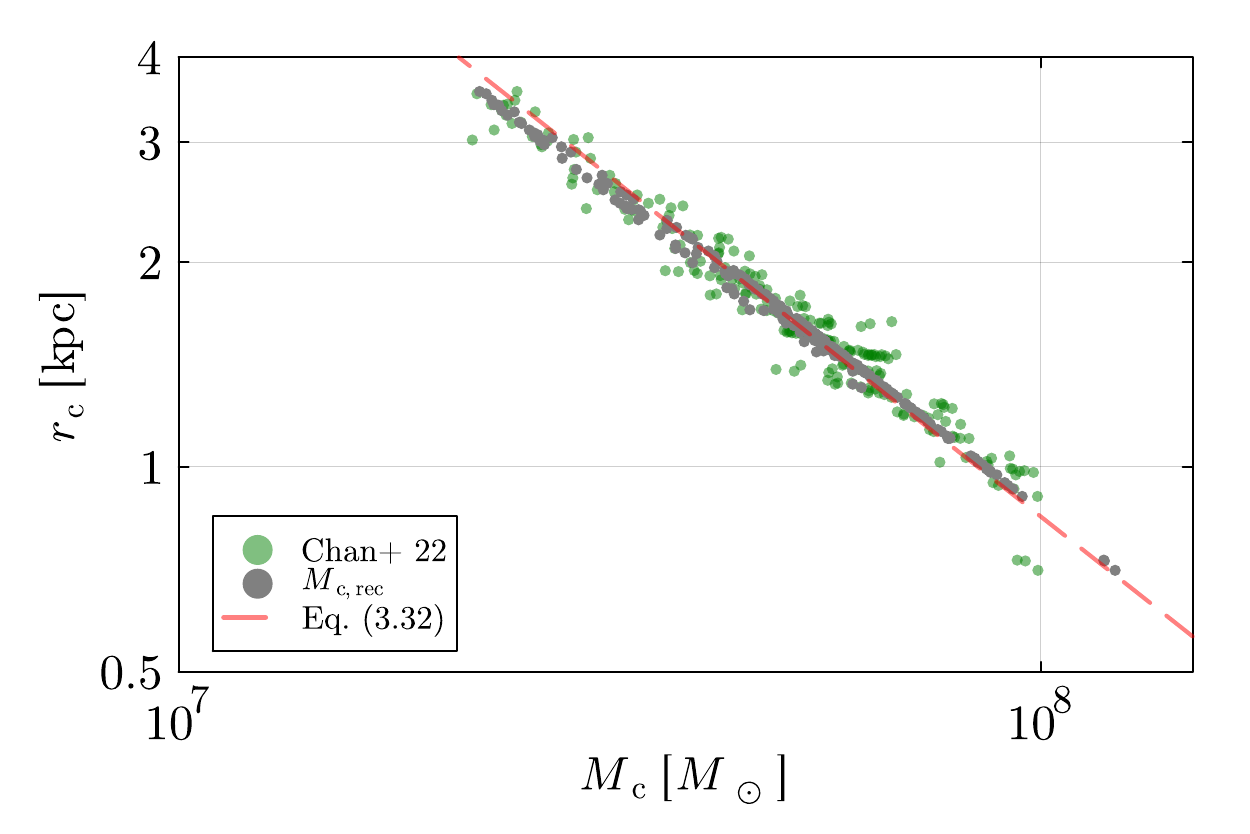}
    \end{minipage}
\caption{
    Relation between core radius and core mass in the cosmological simulation (left)~\cite{May:2021wwp} and the soliton merger simulation (right)~\cite{Chan:2021bja}. The gray points represent the reconstructed core mass values, while the core radius values are taken directly from the simulations. The red dashed line describes the scaling relation eq.~\eqref{eq:mcsol_unit}, which is only valid in the soliton self-gravity dominated region, i.e. the limit A.
}
\label{fig:rc_mc}
\end{figure}

In addition, figure~\ref{fig:rc_mc} illustrates the relationship between core mass and core radius. The purple and green dots represent the core masses obtained from the cosmological and soliton merger simulations, respectively~\cite{May:2021wwp,Chan:2021bja}. The gray dots show the relation between $r_{\rm c}$ and the reconstructed core mass $M_{\rm c,rec}$. Both the simulation and the reconstruction core masses seem to fit well with the scaling relation eq.~\eqref{eq:mcsol_unit} expressed by the red dashed line. 

\acknowledgments
We thank Hei Yin Jowett Chan for sharing the data of the FDM simulation results.
We thank Takahiro Nishimichi, Elisa Ferreira, Neal Dalal, and Hiroki Kawai for their helpful discussion. Research at Perimeter Institute is supported in part by the Government of Canada through the Department of Innovation, Science and Economic Development and by the Province of Ontario through the Ministry of Colleges and Universities. This work was supported by Japan Society for the Promotion of Science (JSPS) Overseas Research Fellowships (Y.M.), JSPS KAKENHI Grant Number JP23KJ1214 (T.T.) and MEXT/JSPS KAKENHI Grants No.~JP20H05861 and No.~JP21H01081 (A.T.).

\bibliographystyle{JHEP}
\bibliography{biblio}
\end{document}